\documentclass[10pt] {article}
\usepackage[usenames,dvipsnames,svgnames,x11names]{xcolor}
\newcommand{\oom}{{\textcolor{Mahogany}{\textsc{oom}}}\xspace}%
\usepackage[pdftex]{graphicx}
\graphicspath{{./figures}}
\usepackage[cmex10]{amsmath}
\usepackage{subfig} 
\usepackage{multirow}
\usepackage{colortbl}
\usepackage{tablefootnote} 
\usepackage[pdftex,colorlinks=true,urlcolor=blue,citecolor=blue]{hyperref}
\usepackage{xspace}
\usepackage{float}
\usepackage{booktabs}

\begin{document}

\title{Understanding the Performance and Power of LLM Inferencing on Edge Accelerators\thanks{Extended version of short paper to appear in PAISE 2025: Mayank Arya and Yogesh Simmhan, ``Understanding the Performance and Power of LLM Inferencing on Edge Accelerators'' in \textit{7th Workshop on Parallel AI and Systems for the Edge, Co-located with IEEE International Parallel \& Distributed Processing Symposium (IPDPS)}, 2025}}

\author{Mayank Arya and Yogesh Simmhan\\
Department of Computational and Data Sciences\\
Indian Institute of Science\\
Bangalore 560012 INDIA\\
Email: \{mayankarya, simmhan\} @iisc.ac.in
}
\date{}

\maketitle

\begin{abstract}

Large Language Models (LLMs) have demonstrated exceptional benefits to a wide range of domains, for tasks as diverse as code generation and robot navigation.
While LLMs are usually served from cloud data centers,
mission critical and privacy sensitive applications may require local hosting of open LLM models. Given the large GPU memory footprint needed for LLMs, edge accelerators such as Nvidia Jetson Orin AGX with 64GB of shared GPU--CPU RAM are a compelling choice. However, the feasibility and performance of LLM inference on edge accelerators is under-explored.
This study presents a detailed evaluation of LLM inference on the NVIDIA Jetson Orin AGX, 
on four SOTA models ranging from $2.7$B to $32.8$B parameters, such as Meta Llama3.1, Microsoft-Phi2, Deepseek-R1-Qwen.We investigate the impact of varying batch sizes, sequence lengths, and quantization levels on latency, throughput and perplexity, and also explore various custom power modes on the Orin AGX to perform power and energy consumption analysis. 
Our findings offer interesting insights on the trade-offs between efficiency, inference speed and resource use, e.g., increasing the sequence length causes a decrease in token throughput and quantization causes smaller LLMs to be slower. These results can help optimize LLM serving on edge accelerators for practical applications.

\end{abstract}

\section{Introduction}

Large Language Models (LLMs) have cemented themselves as the cornerstone of the recent generative Artificial Intelligence (AI) advancements. 
As a specialized subclass of Deep Neural Networks (DNNs), their impressive performance in Natural Language Processing (NLP) tasks, be it text generation, summarization, classification, etc. has made them popular for chatbots, personal virtual assistants, code generation~\cite{chen2021evaluatinglargelanguagemodels}, etc. While the initial phases of the LLM revolution were dominated by closed-source, proprietary models, the landscape is rapidly evolving. An increasing number of models are being released with open access.
Further, model families with different numbers of parameters and tuning are being released to accommodate varying resource footprints and tasks.
As a result, hosting and serving open-source pre-trained and fine-tuned LLMs on local servers is becoming possible. Besides potential cost benefits, this supports latency, privacy and reliability sensitive applications such as robotics and healthcare~\cite{yang2024plug}.
 
\paragraph*{Challenges}
However, the GPU memory footprint even for Small Language Models (SLMs)\footnote{We use the term LLM to refer to language models in general, including small ($<20B$ parameters) and large ($\geq20B$) models.} poses a resource obstacle for local hosting. 

E.g., a server with Nvidia A6000 GPU having 10,752 CUDA cores and 48GB GPU memory -- required to fit LLMs with 20B or more parameters -- and costs over US\$10,000. More recent server GPUs like H100 and H200 are $5$--$10\times$ costlier.

In contrast, high-end edge accelerators like Nvidia Jetson Orin are a viable alternatives for LLM serving. E.g., the latest generation Orin AGX has a 12-core A78AE ARM CPU, an Nvidia Ampere GPU with 2048 CUDA cores, and importantly 64GB of LPDDR5 RAM shared between the GPU and CPU, all in a compact form factor consuming 60W of peak power and costing about US\$2200. These make them a prime candidate for locally hosted LLM inferencing.
This also makes it important to methodically evaluate the performance and limitations for serving LLMs on edge accelerators.

\paragraph*{Related Work and Gaps}
LLM inferencing~ generates output tokens given some input tokens as a prompt. 
Inference is computationally expensive. Model and system optimization techniques such as batching, quantization and caching have also been used to improve its efficiency~\cite{zhou2024survey}.

A recent work~\cite{chittyvenkata2024llminferencebenchinferencebenchmarkinglarge} studies LLM inference benchmarks on a diverse set of server accelerators, including Nvidia A100 and H100, AMD MI300X and MI250, as well as specialized ones like Habana Gaudi2. 

However, edge accelerators were not examined and their insights do not translate directly to Nvidia Jetsons that have shared CPU/GPU memory and 100s of power modes to vary CPU cores, and CPU/GPU/memory frequencies. Others examine Jetson devices for DNN training but omit LLMs~\cite{prashanthi-sigmetrics}.

Seymour et al.~\cite{seymour2024large} also study LLM inferencing performance on Jetson devices. They use the Pythia suite of LLMs and like us evaluate the effect of quantization on latency and energy. However, they consider only the 32GB Orin AGX rather than the more capable 64GB version that we use, and do not run models larger than $1.4B$. Our study uses larger LLMs, and evaluate standard models with $2B$--$32B$ parameters: Microsoft's Phi2 (2.7B), Meta's Llama 3.1 (8B), Mistral's Small 3 (24B) and the recent DeepSeek's R1 Distilled to Qwen (32B) that is supposed to outperform OpenAI o1 mini. 
They also fail to evaluate the effect of batch size, sequence length or diverse power models that can help trade-off performance with energy usage.

Our prior poster has a limited evaluation of older Jetson Xavier AGX 32GB for one workload on small Llama models ($\leq 8B$) but does not examine the effect of sequence length and quantization~\cite{arya2024hipc}, which is key to reducing memory footprint.

\paragraph*{Contributions}
In this short paper, we offer a comprehensive study of the performance and power usage for Orin AGX 64GB on four state-of-the-art (SOTA) LLMs. Specifically, we evaluate the inferencing throughput, latency, power and energy usage for two standard workloads, WikiText2 and LongBench, using different batch sizes and sequence lengths (tokens). We also evaluate their performance and accuracy (perplexity) for quantization of FP32--INT4, e.g., this enables the latest DeepSeek-R1 model with 32B parameter to run on Orin AGX with INT8.
Lastly, we study the effect of 9 different power modes on the performance and power of LLM inferencing.

\section{Experiment Methodology and Setup}

\begin{table}[t]

    \centering    
    \caption{Models used in the experiments. Estimates are shown in \textcolor{red}{red} since the models could not be loaded on device.}
    \renewcommand{\arraystretch}{1.1} 
    \setlength{\tabcolsep}{2pt}
    \small 
    \begin{tabular}{@{}lrrrrr@{}}
        \hline
        \multirow{2}{*}{\textbf{Model Name (HF)}} & 
        \multirow{2}{*}{\textbf{\# Params}} &         
        \multicolumn{4}{c}{\textbf{Peak Memory (GB)}}\\
        &&\textbf{FP32} & \textbf{FP16} & \textbf{INT8} & \textbf{INT4}\\
        \hline\hline
        Microsoft Phi-2\tablefootnote{\url{https://huggingface.co/microsoft/phi-2}}  & 2.7B & 11.2 & 5.6  & 3.0  & 1.8\\
        Meta Llama-3.1-8B\tablefootnote{\url{https://huggingface.co/meta-llama/Llama-3.1-8B}} & 8.0B & 32.2 & 16.1  & 9.1  & 5.6  \\
        Mistral-Small-24B\tablefootnote{\url{https://huggingface.co/mistralai/Mistral-Small-24B-Base-2501}} & 23.6B & \textcolor{red}{${\approx 94.2}$} & 47.1  & 24.9  & 13.8  \\
        DeepSeek-R1-Qwen-32B\tablefootnote{\url{https://huggingface.co/deepseek-ai/DeepSeek-R1-Distill-Qwen-32B}} & 32.8B & \textcolor{red}{${\approx 124.0}$} & \textcolor{red}{${\approx 62.0}$}  & 34.3  & 18.7  \\
        \hline
    \end{tabular}
    \label{tab:llm_models}
   
\end{table}

\paragraph{Language Models and Workloads}
Table~\ref{tab:llm_models} lists the four SOTA LLMs with parameter counts from $2$--$32B$ that we evaluate. 

This includes \textit{Microsoft Phi-2}, a compact transformer optimized for efficiency with code and reasoning capabilities; \textit{Llama 3.1}, Meta's latest open-weight model for general-purpose tasks; \textit{Mistral-Base}, a high-capacity dense model for reasoning and text generation; and \textit{DeepSeek-Qwen-Distil}, dense model distilled from the recent DeepSeek-R1 based on Qwen. This diverse set helps us analyze inference behavior across different LLM families and target domains.

We generate prompts to the LLMs from WikiText2~\cite{merity2016pointer} and LongBench~\cite{bai2023longbench}. We extract paragraphs with $\geq 256$ tokens as a pool of valid prompts. For each inference batch, we randomly sample the required number of prompts. 

\paragraph{Edge Hardware}
We conduct our experiments on the NVIDIA Jetson Orin AGX Developer Kit (64GB). It has a 12-core ARM A78AE CPU@$2.2$GHz, an Ampere GPU with $2048$ CUDA cores@$1.3$ GHz, and $64$GB of LPDDR5 RAM shared between CPU and GPU. 
The Orin AGX is flashed with JetPack v$6.0$ with CUDA v$12.2$, and runs Ubuntu $22.04$ LTS. We use PyTorch v$2.3.0$ and Hugging Face Transformers library for inferencing. All datasets and models are loaded from SSD.

\paragraph{Experiment Configurations}

Since GPUs are optimized for parallel computation, LLM inference workloads often achieve higher efficiency and throughput with larger \textit{batch sizes}, increasing the number of prompts inferenced together. However, increasing the batch size ($bs$) also drives up the inference latency per prompt. We vary the batch size from $bs=1$ to $128$, in powers of $2$, and examine the extent to which throughput benefits are seen and the latency tradeoff. By default, we use a batch size of $32$.

LLM models consume and produce tokens, and there is a growing trend to increase the number of tokens processed at a time. The total number number of tokens present in the input prompt and output response together is called the \textit{sequence length (sl)}.

Increasing the sequence length directly affects the computational complexity, memory usage and latency per prompt.
We vary sequence length from $sl=128$--$1024$ tokens, in powers of $2$, and study the effect of short and long prompts and responses on the performance.
We use a diverse subset or multiples of the 256-token prompts to form a single input, and limit the output tokens to the remaining sequence length.

\textit{Quantization} refers to the use of lower precision arithmetic to perform model inferencing rather than full 32-bit floating point (FP32). Using quantization allows the model weights loaded into memory to use a lower precision, leading to a smaller memory footprint and allowing larger models to be used. However, this can reduce the accuracy of the outputs and also leads to higher computing overheads, of 14--86\% seen for GPT models, due to inefficient saturation of GPU~\cite{NEURIPS2022_c3ba4962}.
We use the popular \textit{BitsAndBytes LLM.int8()} tool~\cite{NEURIPS2022_c3ba4962} to quantize the models to three lower precisions: FP16, INT8 and INT4.

\begin{table}[t]

    \centering
    \renewcommand{\arraystretch}{1}
    \caption{Resource configurations for power modes evaluated. Highlighted resource varies between rows.}
    \begin{tabular}{l c c c c}
        \hline
        \bf Power Mode & \begin{tabular}[c]{@{}c@{}}\bf GPU Freq \\ \bf (MHz)\end{tabular} & \begin{tabular}[c]{@{}c@{}}\bf CPU Freq \\ \bf (GHz)\end{tabular} & \begin{tabular}[c]{@{}c@{}}\bf CPU Cores \\ \bf Online\end{tabular} & \begin{tabular}[c]{@{}c@{}}\bf Memory \\ \bf Freq (GHz)\end{tabular} \\
        \hline\hline
        MaxN & $1301$ & $2.2$ & $12$ & $3200$ \\
        \hline
        A & \cellcolor{blue!15} \textbf{$800$}  & $2.2$ & $12$ & $3200$ \\
        B & \cellcolor{blue!15} \textbf{$400$}  & $2.2$ & $12$ & $3200$ \\
        \hline
        C & $1301$ & \cellcolor{blue!15} \textbf{$1.7$} & $12$ & $3200$ \\
        D & $1301$ & \cellcolor{blue!15} \textbf{$1.2$} & $12$ & $3200$ \\
        \hline
        E & $1301$ & $2.2$ & \cellcolor{blue!15} \textbf{$8$}  & $3200$ \\
        F & $1301$ & $2.2$ & \cellcolor{blue!15} \textbf{$4$}  & $3200$ \\
        \hline
        G & $1301$ & $2.2$ & $12$ & \cellcolor{blue!15} \textbf{$2133$} \\
        H & $1301$ & $2.2$ & $12$ & \cellcolor{blue!15} \textbf{$665$} \\
        \hline
    \end{tabular}
    \label{tab:power_modes}

\end{table}

Orin AGX, like other Jetson devices, offers 1000s of \textit{power modes (PM)}, with MAXN being the default and fastest.
Custom power modes control the CPU, GPU and memory frequencies, and set the number of active CPU cores. We study eight custom power modes that vary these dimensions (Table~\ref{tab:power_modes}) to identify trade-offs between power and performance.

\paragraph{Evaluation Metrics}
\label{sec:eval-metric}
 
    We measure \textit{Token Throughput ($tokens/s$)} as the number of input and output tokens processed per second in a batch, given by $TP = \sum_{i \in Batch} \frac{(\#Input\ tokens_i + \#Output\ tokens_i)}{Batch\ Latency}$. This determines the inference efficiency when processing concurrent requests. 
    We report \textit{latency ($s$)}, as the end-to-end time taken to generate all output tokens for all prompts in a given batch, time to last token for the batch.

    Orin AGX has memory shared between CPU and CPU. We measure and report the \textit{incremental peak memory usage ($GB$)} for each inferencing workload, which is the difference between the peak memory usage during the run and the base memory usage before loading the model. We occasionally report the memory used to load the model, before the workload starts. \oom indicates out of memory. 
    We log the system power metrics using \textit{jtop}
    and report the \textit{median power usage ($Watts$)} across batches.  
    For \textit{total energy usage}, we perform trapezoidal numerical integration over time for a batch with power sampled every $2$s, and summed across all batches.

    We evaluate the impact of quantization on model accuracy using \textit{perplexity scores}, given by $\exp\left(\frac{\sum \text{NLL}}{\text{total tokens}}\right)$, where \( \sum \text{NLL} \) is the total negative log-likelihood over all valid tokens. It measures how well a model predicts a sequence of text, with \textit{lower values indicating better performance}. 
    For WikiText2 and LongBench, we process text in overlapping windows of 1024 tokens with a stride of 512. The model's loss, computed using cross-entropy, represents the negative log-likelihood of the target tokens (Table \ref{tab:ppl_scores}).

To ensure reliable and stable measurements, we conduct a warm-up run to mitigate initialization overhead, followed by five actual runs for each configuration, averaging the results across these runs. This approach minimizes variability and provides a more accurate assessment of inference performance.  

\section{Results and Analysis}
\subsection{Impact of Batch Size}
\label{subsec:batch-size-impact}

\begin{figure}[t]

    \centering
    \includegraphics[width=0.9\linewidth]{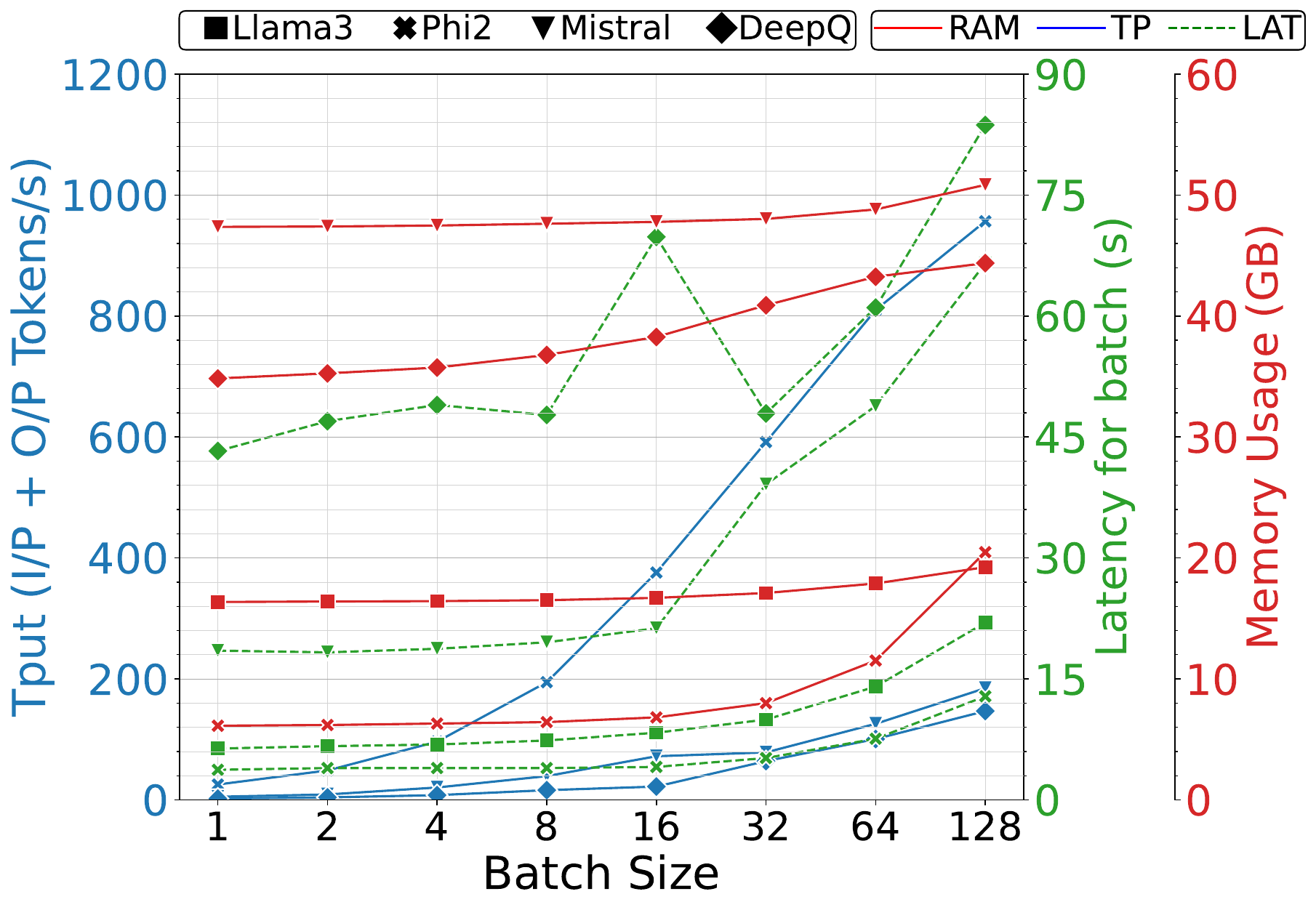}
   
    \caption{Varying \textit{batch size} across models. MaxN with $sl=96$ and FP16 (INT8 for DeepQ).}
    \label{fig:vary-batch-size}
   
\end{figure}

Fig.~\ref{fig:vary-batch-size} shows the effect of increasing batch size on the throughput (blue solid line, left Y axis) and latency (green dashed line, right inner axis). While we report results for wikitext2 for brevity, all trends hold for LongBench as well. 

As expected, increasing the batch size leads to a higher throughput as more tokens are processed in parallel, e.g., improving by $203\%$ from $184$ to $558$~tok/s for Llama as $bs$ grows from $32$ to $128$. 

However, this also results in increased latency, i.e., the time to last token for prompts in the batched is delayed, with latency increasing from $10$ to $22s$ for Llama3 from $bs=32$ to $128$. 

The largest Deepseek model starts to saturate concurrency and throughput at $bs=128$.

The incremental memory utilization also increases with batch size, as the prompts being processed increases. This reaches a peak of $57.2$ GB for Deepseek at $bs=128$, with the model memory being $47.1$GB. This growth is due to higher usage of key-value (KV) cache used to store previously computed key and value tensors for the attention mechanisms. This grows linearly with the total sequence count for the batch. 

So, while larger batch sizes improve GPU efficiency, they also push memory constraints.

\subsection{Impact of Sequence Length}
\label{subsec:impact-of-seqlen}
\begin{figure}[t]
    \centering
    \includegraphics[width=0.9\linewidth]{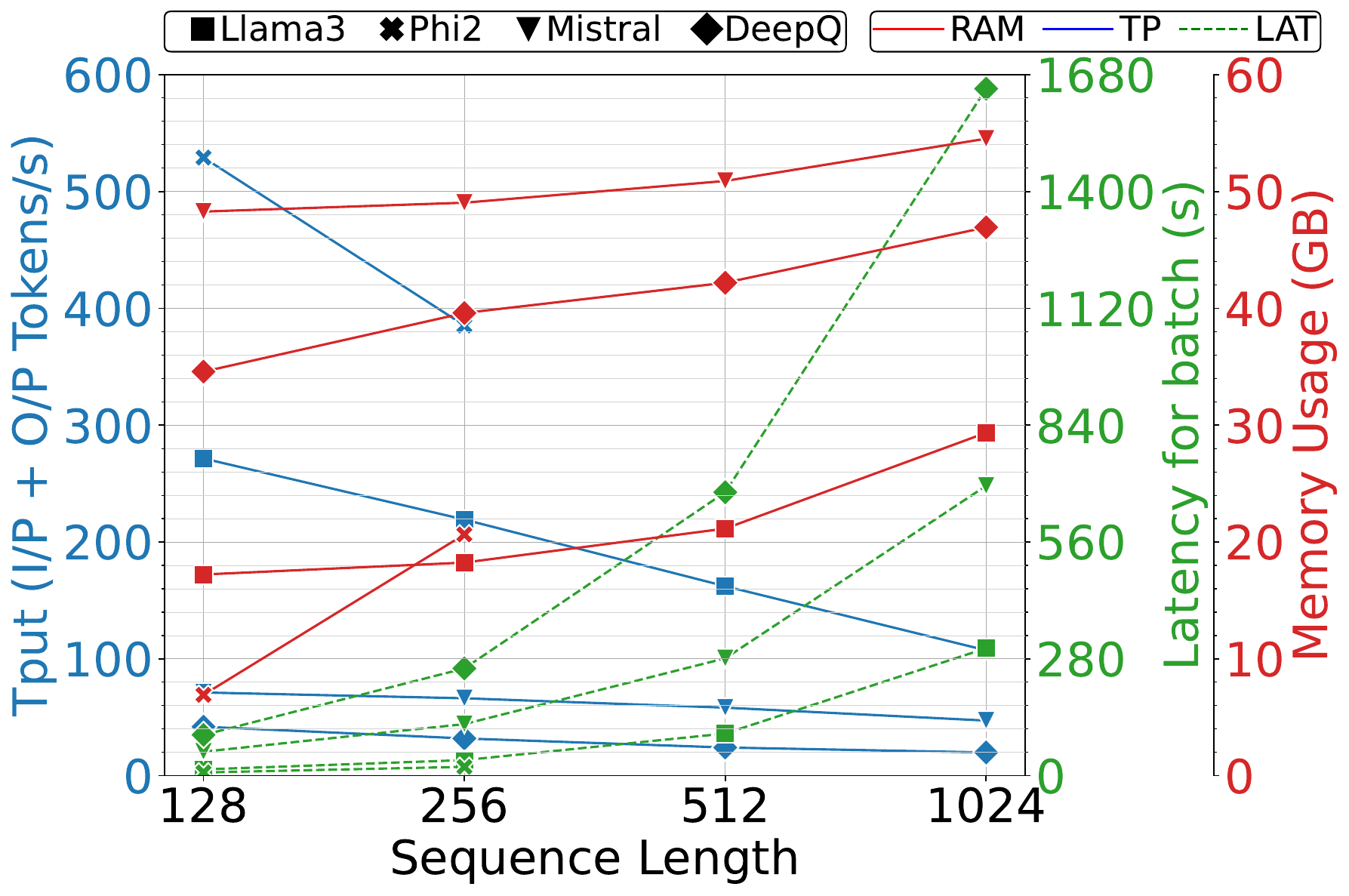}
    
    \caption{Varying \textit{sequence length} across models. MAXN with $bs=32$ and FP16 (INT8 for DeepQ). Phi2 is OOM for sequence length $> 256$.}
    \label{fig:vary-seq-len}
   
\end{figure}

    Fig.~\ref{fig:vary-seq-len} shows that as we increase the sequence length, the throughput decreases. This is due to how LLMs generate text and the way we structure our sequence lengths. In our experiments, inference is dominated by the auto-regressive \textit{decode phase} (output tokens being generated sequentially). Specifically, we evaluate sequence lengths defined as A = B + C, where B is the number of input tokens and C is the number of generated tokens. The configurations we consider are: $\textbf{128}$ ($32$ + $96$), $\textbf{256}$ ($64$ + $192$), $\textbf{512}$ ($128$ + $384$), and $\textbf{1024}$ ($256$ + $768$). In the decode phase, one token is generated per iteration through the model, where the token must refer to the context of all previous tokens generated before it. This dependency makes the decode phase \textit{memory bound}~\cite{patel2024splitwiseefficientgenerativellm}.
 
    Since the batch size here is fixed at $bs=32$ limiting further parallelism, longer sequence lengths cause the throughput to drop.
    E.g., as length increases from $128$ to $1024$, Llama3.1's throughput drops from $271$ to $107tok/s$ as the latency grows from $15s$ to $305s$.

    A larger sequence length also leads to more memory usage as the tokens stored in the KV cache grows, increasing from $17.2$GB to $29.4$GB in the above example, even though the base model takes only $16.1$ GB of memory.

\subsection{Impact of Quantization}
\label{subsec:impact-of-quantization}
\begin{figure}[t]

    \centering
    \includegraphics[width=0.95\linewidth]{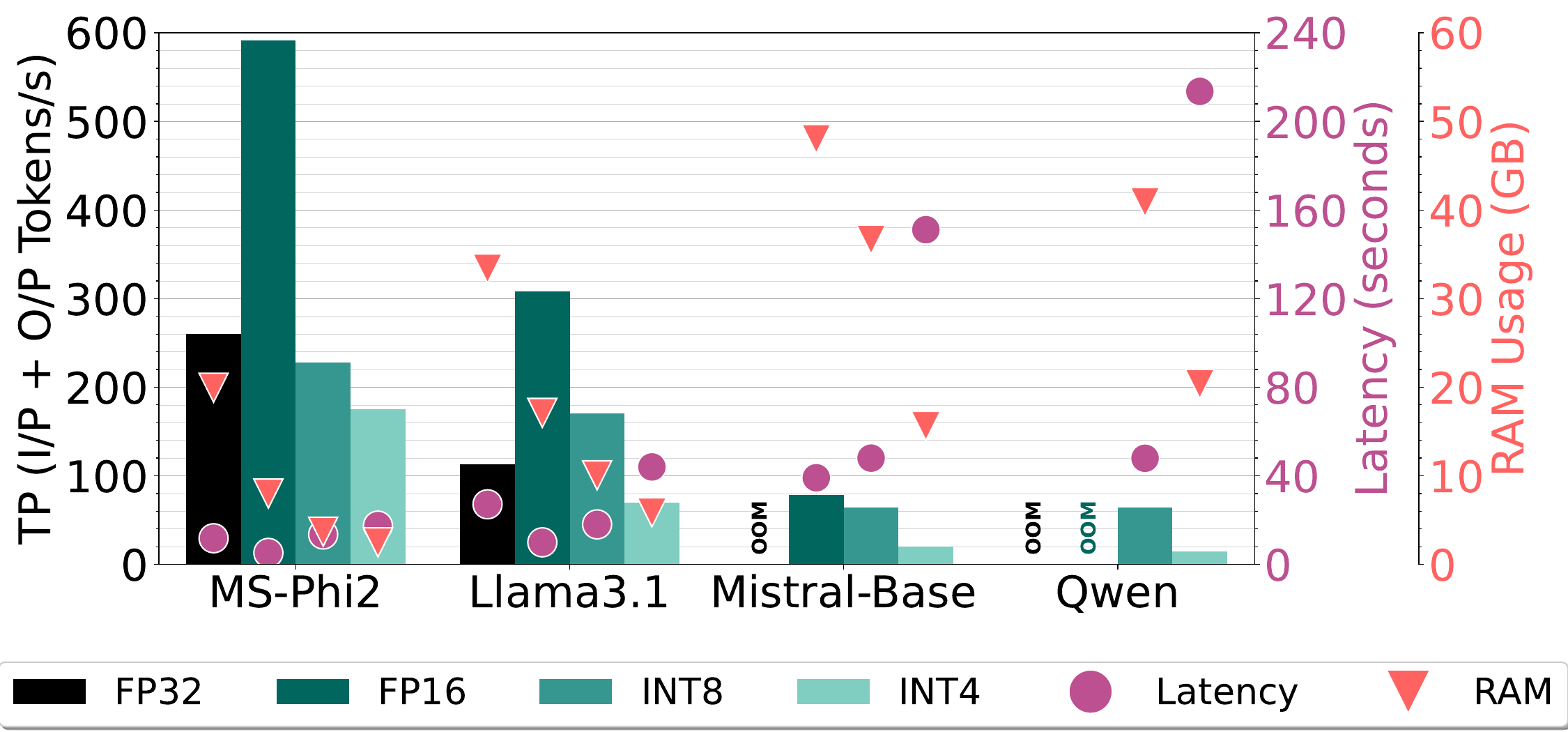}
    \caption{Impact of \textit{quantization} across models. MAXN with $bs=32$ and $sl=96$. OOM stands for out-of-memory.}
    \label{fig:vary-quantization}
    
\end{figure}

\begin{table}[t]

    \centering
    \setlength{\tabcolsep}{2.5pt} 
    \renewcommand{\arraystretch}{1.1} 
    \caption{\textit{Perplexity} of models with different precision.  }
      \begin{tabular}{l|cccc|cccc}
        \hline
        \multirow{2}{*}{\bf Model} & \multicolumn{4}{c|}{\bf WikiText2} & \multicolumn{4}{c}{\bf LongBench} \\
        \cline{2-9}
        & \bf FP32 & \bf FP16 & \bf INT8 & \bf INT4 & \bf FP32 & \bf FP16 & \bf INT8 & \bf INT4 \\
        \hline\hline
        MS-PHI2 & $9.12$ & $9.12$ & $9.34$ & $9.69$ & $7.35$ & $7.35$ & $7.47$ & $7.65$ \\
        Llama3 & $5.91$ & $5.91$ & $6.00$ & $6.30$ & $5.77 $& $5.77$ & $5.80$ & $5.99$ \\
        Mistral-Base & \oom & $4.99$ & $5.00$ & $5.08$ & \oom & $4.95$ & $4.97$ & $5.11$ \\
        Deepseek-Qwen & \oom & \oom & $6.36$ & $6.48$ & \oom & \oom & $6.42$ & $6.53$ \\
        \hline
    \end{tabular}
    \label{tab:ppl_scores}
 
\end{table}

The memory efficiency of quantization comes at the cost of not just lower accuracy but also increased latency, especially for smaller models with low hidden dimensions~\cite{NEURIPS2022_c3ba4962}. 
As Fig.~\ref{fig:vary-quantization} shows, for smaller models such as Phi-2 and Llama3.1-8B, INT8 inference reduces RAM usage by about $46\%$ but is slower by $62\%$ than FP16.
Quantization introduces compute overhead for dequantization and quantization-aware processing which disproportionately impacts smaller models with limited parallel compute utilization.
For the larger Mistral-Base-24B, INT8 is within $2\%$ of FP16 latency, offering lower penalties, and still reducing RAM by $47\%$.

These trends are unlike those seen on Nvidia A100, where they observe faster performance upon quantization for models $>13B$~\cite{NEURIPS2022_c3ba4962}.

As expected, perplexity (accuracy) becomes worse with quantization (Table~\ref{tab:ppl_scores}). This is marginal from FP16 to INT8, rising from $5.9$ to $6.0$ from for Llama and $4.99$ to $5$ for Mistral, but sharper from INT8 to INT4, growing to $6.30$ and $5.08$ for these LLMs.
 
\begin{figure}[t]
    \centering
    \includegraphics[width=0.9\linewidth]{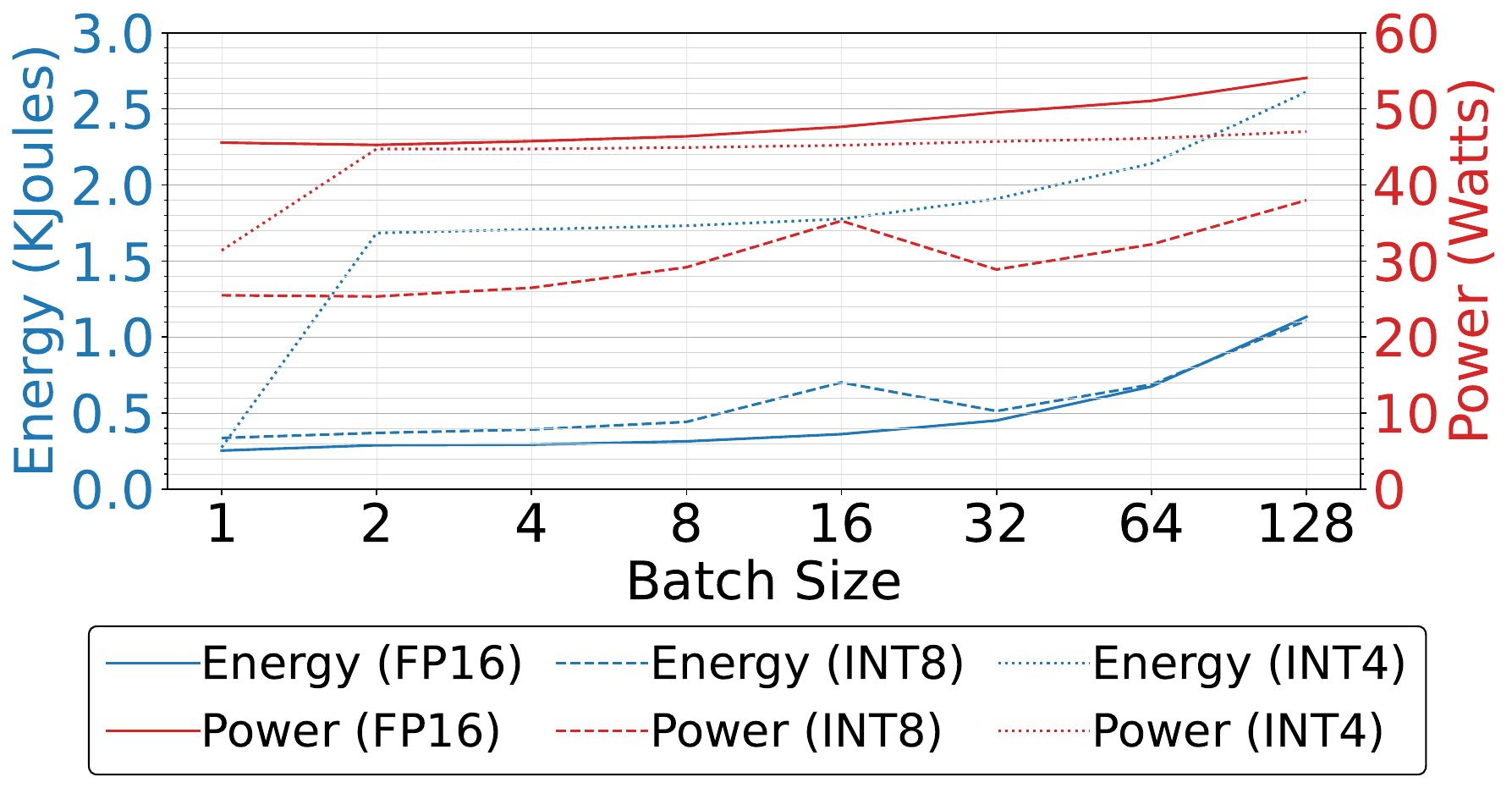}
  
    \caption{\textit{Power load and Energy use} when varying \textit{batch sizes and quantization} for Llama3.1. MAXN with $sl=96$.}
    \label{fig:L3-power-energy}
\end{figure}

Fig \ref{fig:L3-power-energy} 

shows that quantization worsens energy usage due to higher latency. Notably, INT8's energy use is comparable to FP16 at higher batch sizes, while INT4's energy use is higher due to latency degradation. Our experiments also indicate that INT8 uses only $\approx 60\%$ of the GPU while INT4 uses $100\%$, which can contribute to its increased power draw. This opens up opportunities for further investigation of memory--latency--energy trade-offs due to quantization.

\subsection{Impact of Power Modes}

\begin{figure}

    \centering
    \includegraphics[width=0.95\linewidth]{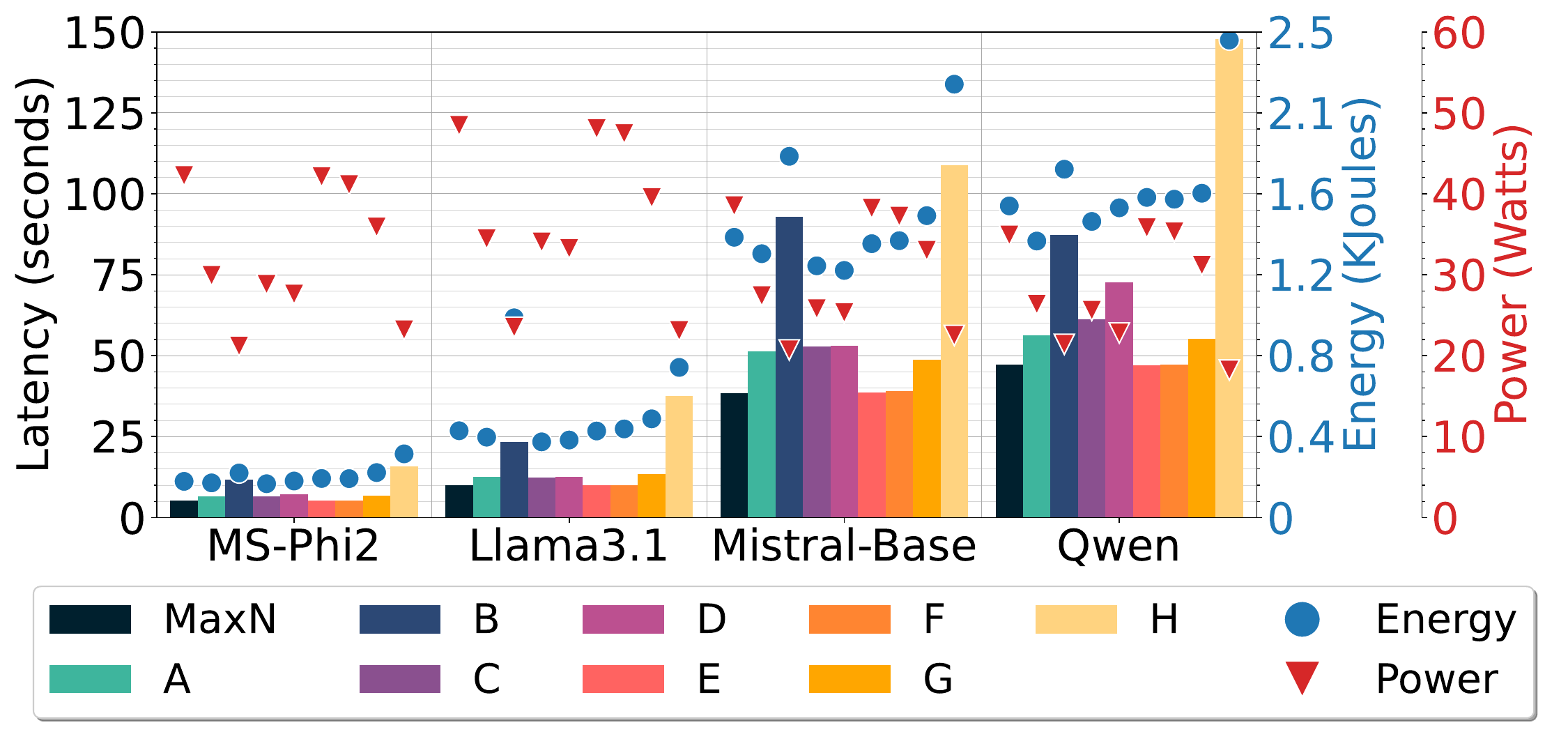}

    \caption{\textit{Varying power modes} across models. $bs=32$, $sl=96$ and FP16 (INT8 for DeepQ). Bars are latency (left y-axis) and markers are energy use and power (right y-axes).}
    \label{fig:power-energy-consumption}
  
\end{figure}

We focus on Llama for the analysis of various power modes (Fig.~\ref{fig:power-energy-consumption}), but see similar trends for others. Power Mode (PM) \textbf{A} which has a lower GPU speed than MAXN reduces instantaneous power by $\approx28\%$ and also leads to lower overall energy consumption since there is a smaller increase in inference latency $\approx 26\%$ for the workload. \textbf{PM-B} further lower the GPU speed and power reduces by $51\%$ over MAXN. But this has a higher impact on latency and leads to an increase in total energy usage than MAXN.
So this is better suited for power-constrained setups rather than for energy mitigation. 

Limiting CPU frequency in \textbf{PM-C} and \textbf{PM-D} reduces power by $30\%$ and increases latency by $25\%$. 
Interestingly, \textbf{PM-A} (GPU freq. lower by $38\%$) and \textbf{PM-C} (CPU freq. lower by $22\%$) have a similar impact on latency relative to MAXN for a given model, e.g., causing Phi-2 to slow down by 1.3\% in both cases and Mistral by 14\%, 
except \textit{Deepseek} which runs in INT8 is likely using CPU to assist with quantization.

In contrast, reducing the CPU core-count in \textbf{PM-E} and \textbf{PM-F} has negligible latency impact compared to \textbf{MaxN}, suggesting limited core-parallelism.

Finally, reducing the memory frequency has a significant effect; in \textbf{PM-H}, latency increases by $370\%$ over MAXN and energy usage by $72\%$ even as power load drops by $52\%$. 

\section{Conclusions}

In this study, we explore LLM inferencing on the Jetson Orin 64GB edge accelerator using SOTA models on two workloads.
We offer the first systematic investigation of the impact of key system, workload and model optimization factors such as batch size, sequence length, quantization and power modes on performance and quality metrics for this edge platform. 

Our findings confirm the viability of modern edge accelerators to support LLM serving, though challenges remain in enhancing latency and energy usage.
Future work can further analyze these empirical results, and leverage them to optimize LLM inferencing on the edge, including the use of dedicated inference engines, accelerators like DLAs, coupling edge inferencing with cloud endpoints, etc. ultimately broadening the accessibility of LLMs for real-world applications.
\section{Acknowledgements}
 We thank students from the DREAM:Lab, IISc, and in particular Prashanthi S.K., for their guidance and assistance on this paper.

\bibliographystyle{IEEEtran}
\bibliography{arxiv}

\begin{thebibliography}{10}
\providecommand{\url}[1]{#1}
\csname url@samestyle\endcsname
\providecommand{\newblock}{\relax}
\providecommand{\bibinfo}[2]{#2}
\providecommand{\BIBentrySTDinterwordspacing}{\spaceskip=0pt\relax}
\providecommand{\BIBentryALTinterwordstretchfactor}{4}
\providecommand{\BIBentryALTinterwordspacing}{\spaceskip=\fontdimen2\font plus
\BIBentryALTinterwordstretchfactor\fontdimen3\font minus \fontdimen4\font\relax}
\providecommand{\BIBforeignlanguage}[2]{{%
\expandafter\ifx\csname l@#1\endcsname\relax
\typeout{** WARNING: IEEEtran.bst: No hyphenation pattern has been}%
\typeout{** loaded for the language `#1'. Using the pattern for}%
\typeout{** the default language instead.}%
\else
\language=\csname l@#1\endcsname
\fi
#2}}
\providecommand{\BIBdecl}{\relax}
\BIBdecl

\bibitem{chen2021evaluatinglargelanguagemodels}
M.~Chen \emph{et~al.}, ``Evaluating large language models trained on code,'' arXiv, Tech. Rep. 2107.03374, 2021.

\bibitem{yang2024plug}
Z.~Yang, S.~S. Raman, A.~Shah, and S.~Tellex, ``Plug in the safety chip: Enforcing constraints for llm-driven robot agents,'' in \emph{IEEE ICRA}, 2024.

\bibitem{zhou2024survey}
Z.~Zhou \emph{et~al.}, ``A survey on efficient inference for large language models,'' arXiv, Tech. Rep. 2404.14294, 2024.

\bibitem{chittyvenkata2024llminferencebenchinferencebenchmarkinglarge}
K.~T. Chitty-Venkata \emph{et~al.}, ``Llm-inference-bench,'' arXiv, Tech. Rep. 2411.00136, 2024.

\bibitem{prashanthi-sigmetrics}
{S. K., Prashanthi}, S.~A. Kesanapalli, and Y.~Simmhan, ``Characterizing the performance of accelerated jetson edge devices for training dnns,'' in \emph{SIGMETRICS}, 2023.

\bibitem{seymour2024large}
L.~Seymour, B.~Kutukcu, and S.~Baidya, ``Large language models on small resource-constrained systems: Performance characterization, analysis and trade-offs,'' arXiv, Tech. Rep. 2412.15352, 2024.

\bibitem{arya2024hipc}
M.~Arya and Y.~Simmhan, ``A preliminary performance analysis of llm inference on edge accelerators,'' in \emph{IEEE HiPCW}, 2024.

\bibitem{merity2016pointer}
``Wikitext-2,'' 2016, \url{https://paperswithcode.com/dataset/wikitext-2}.

\bibitem{bai2023longbench}
``Longbench,'' 2023, \url{https://github.com/THUDM/LongBench}.

\bibitem{NEURIPS2022_c3ba4962}
T.~Dettmers, M.~Lewis, Y.~Belkada, and L.~Zettlemoyer, ``Gpt3.int8(): 8-bit matrix multiplication for transformers at scale,'' in \emph{NeurIPS}, 2022.

\bibitem{patel2024splitwiseefficientgenerativellm}
P.~Patel \emph{et~al.}, ``Splitwise: Efficient generative llm inference using phase splitting,'' arXiv, Tech. Rep. 2311.18677, 2024.

\end{thebibliography}

\clearpage

\appendix
\section{Appendix}
These additional experiments and results complement the original results published in PAISE 2025.

\subsection{Impact of Varying Batch Size}

\begin{figure}[h]
    \centering
    \subfloat[Deepseek-Qwen-Distill\label{fig:deepQ-vary-batch}]{
        \includegraphics[width=0.48\textwidth]{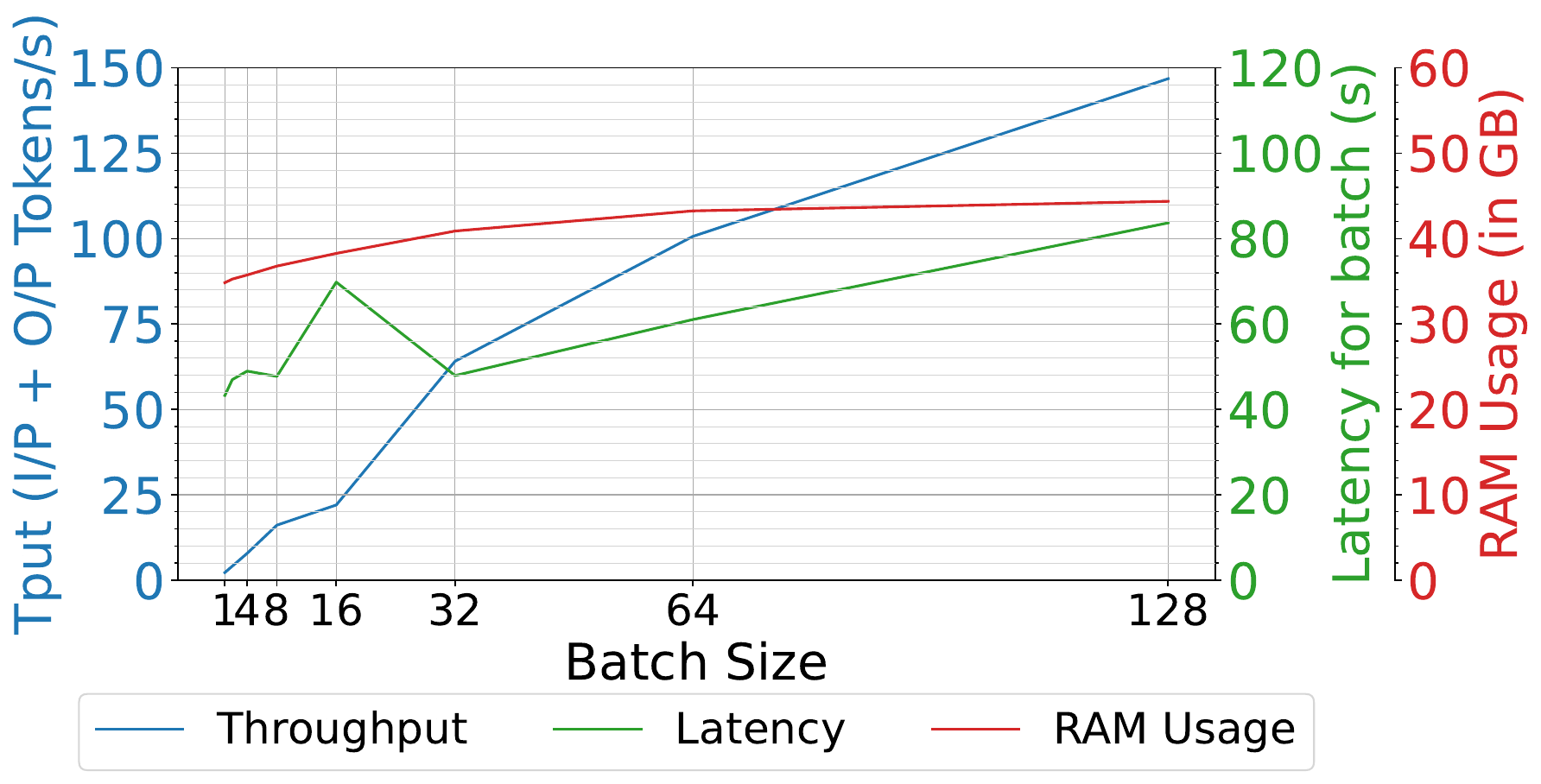}
    }
    \hfill
    \subfloat[Llama3.1-8B\label{fig:L3-vary-batch}]{
        \includegraphics[width=0.48\textwidth]{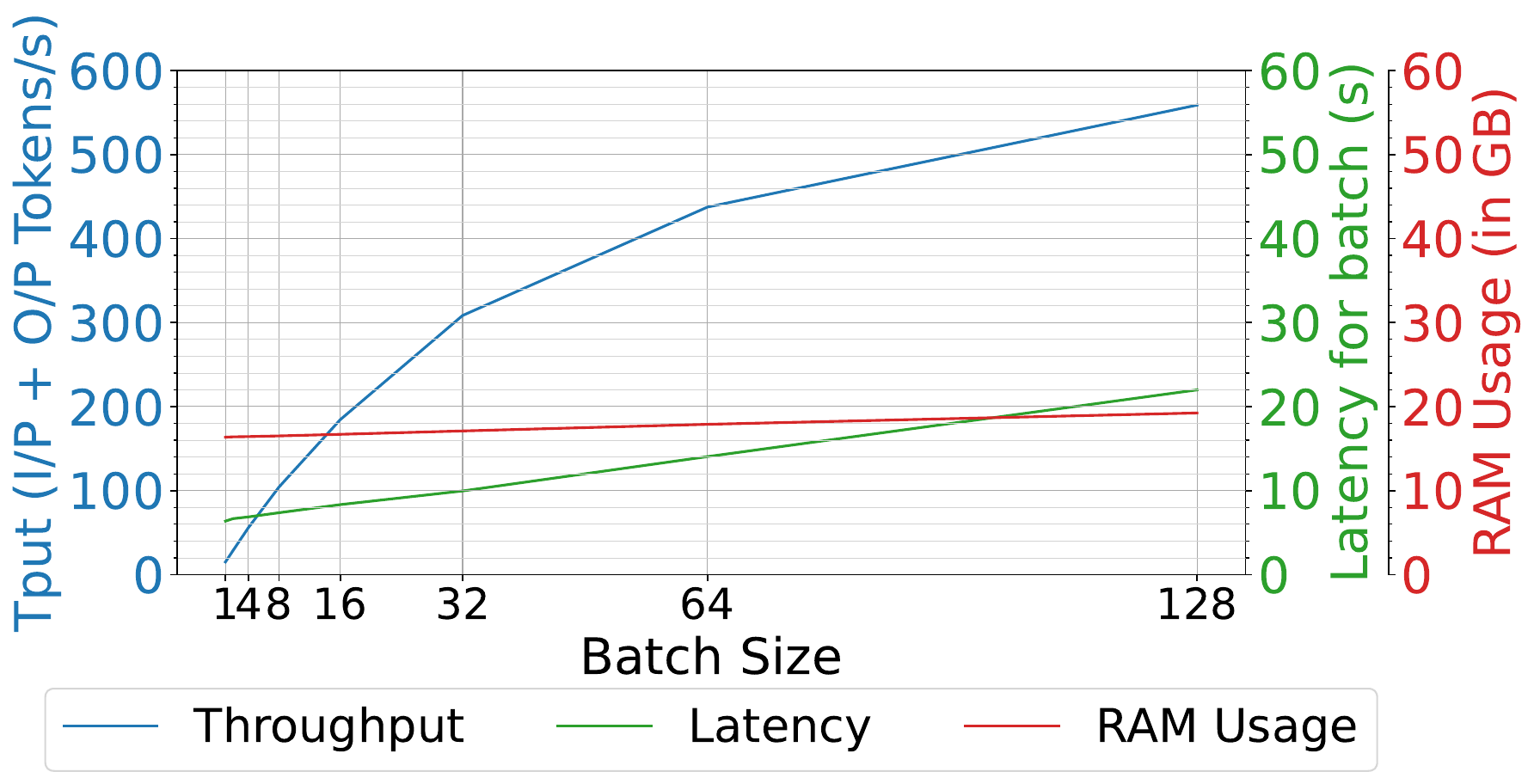}
    }
    
    \vspace{0.5cm}  
    
    \subfloat[Mistral-Base\label{fig:mistral-vary-batch}]{
        \includegraphics[width=0.48\textwidth]{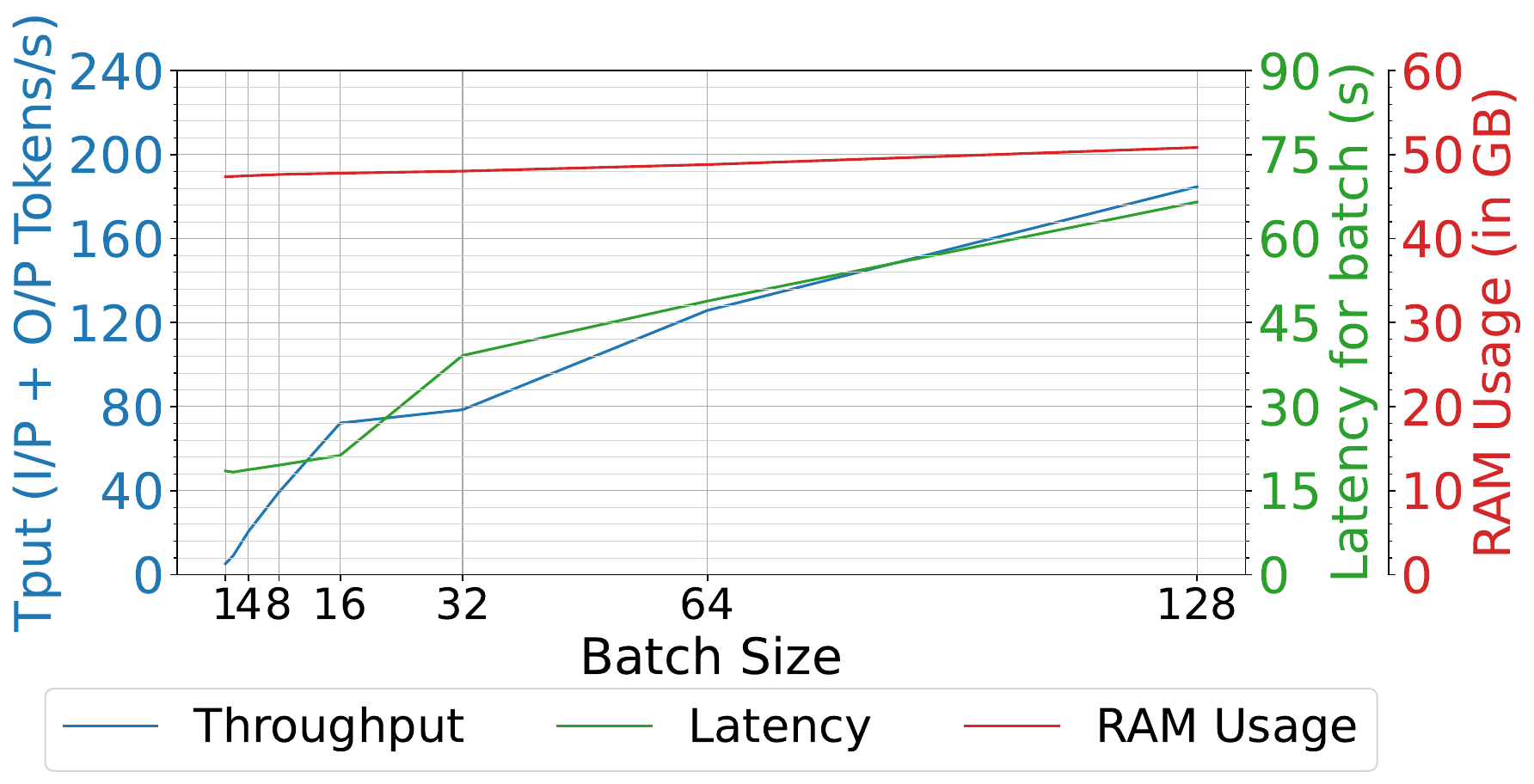}
    }
    \hfill
    \subfloat[MS-Phi2\label{fig:MS-Phi2-vary-batch}]{
        \includegraphics[width=0.48\textwidth]{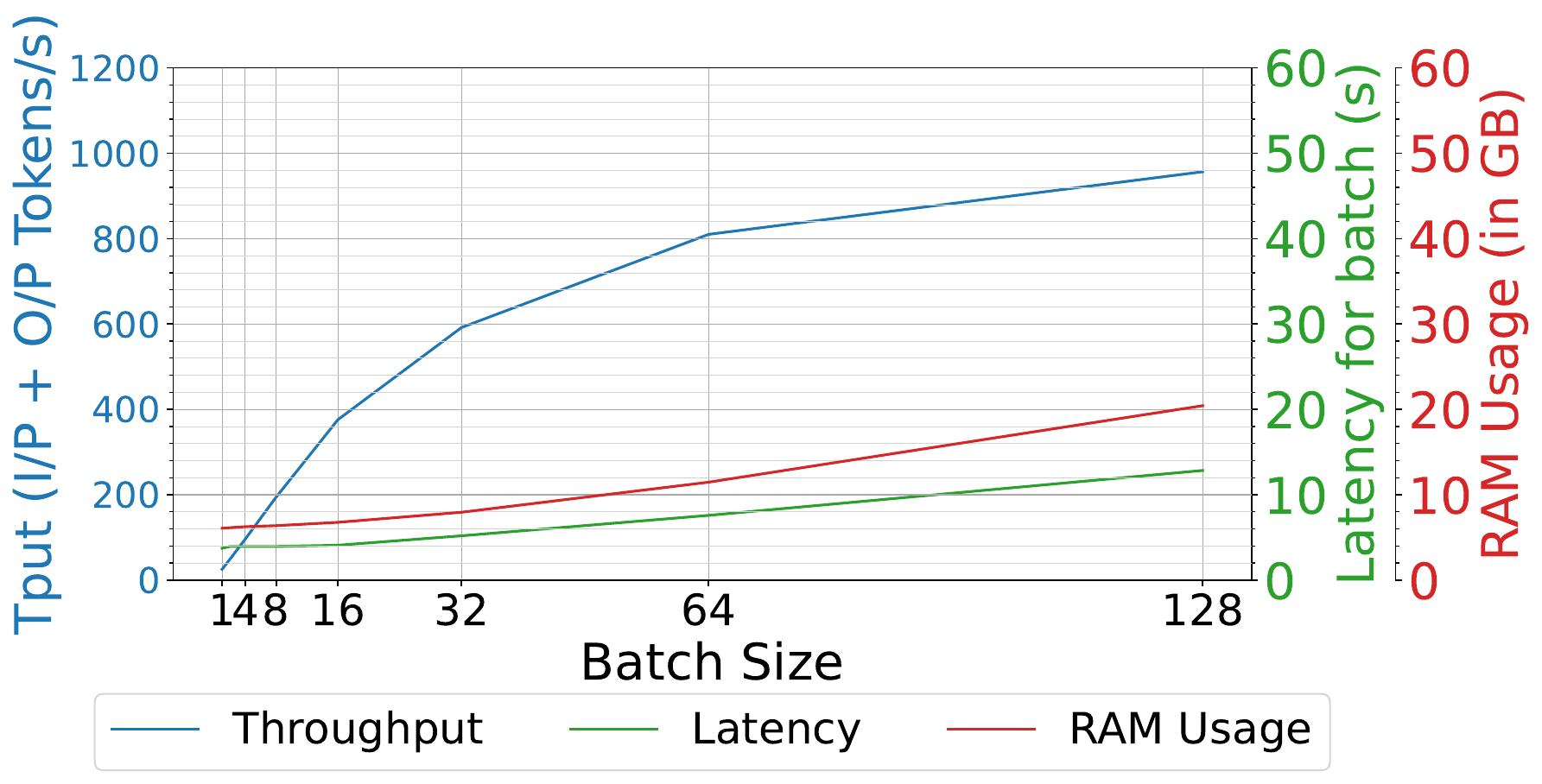}
    }
    
    \caption{Impact of varying batch sizes on different models evaluated using the \textbf{WikiText2} dataset. All models operate in \textbf{FP16} precision, except \textbf{Deepseek-Qwen}, which uses \textbf{INT8}. Experiments were conducted with Power Mode set to \textbf{MaxN} and sequence length (seqlen) equal to \textbf{96}, comprising 32 input tokens and 64 generated tokens.}
    \label{fig:varying-batch-size-detailed}
\end{figure}

Figure \ref{fig:varying-batch-size-detailed} shows the impact of increasing batch sizes (\S\ref{subsec:batch-size-impact}) on the memory consumption, end-to-end latency, and throughput for the \textit{WikiText2} dataset of various models evaluated in this study in detail. As the batch size grows, there is a predictable rise in total memory usage, encapsulating both the model weights and the memory allocated for intermediate results, such as the key-value (KV) cache. For instance, the MS-Phi2 model requires approximately $5.6$ GB to load its weights into memory. With a batch size of $1$, it consumes an additional $0.58$GB, totaling $6.18$ GB. When the batch size is increased to $128$, the memory usage for processing rises to $14.93$ GB, leading to a cumulative memory consumption of $20.53$ GB. The throughput increases from $25$ to $956$ tokens per sec, while the end-to-end latency for the batch increases from $3.73$ to $12.85$ seconds. Similarly Table \ref{tab:vary-batch-WT} shows the performance of all the other models on the \textit{WikiText2} dataset.

\begin{table}[H]
    \centering
    \resizebox{\textwidth}{!}{ 
    \begin{tabular}{c|cccc|cccc|cccc}
        \toprule
        \textbf{Batch Size} & \multicolumn{4}{c|}{\textbf{RAM (GB)}} & \multicolumn{4}{c|}{\textbf{Latency (ms)}} & \multicolumn{4}{c}{\textbf{Throughput (tokens/s)}} \\
        & Phi2 & Llama3 & Mistral & DeepQ & Phi2 & Llama3 & Mistral & DeepQ & Phi2 & Llama3 & Mistral & DeepQ \\
        \midrule
        1   & 6.18
& 16.38
& 47.33
& 34.82
& 3.73
& 6.37
& 18.51
& 43.25
& 25.45
& 15.08
& 5.19
& 2.22
\\
        2   & 6.24
& 16.42
& 47.36
& 35.24
& 3.95
& 6.66
& 18.3
& 46.97
& 48.66
& 28.82
& 8.96
& 4.09
\\
        4   & 6.36
& 16.45
& 47.44
& 35.72
& 3.95
& 6.87
& 18.74
& 48.97
& 96.24
& 55.91
& 20.49
& 7.84
\\
        8   & 6.48
& 16.53
& 47.59
& 36.76
& 3.95
& 7.37
& 19.54
& 47.73
& 194.59
& 104.27
& 39.3
& 16.09
\\
        16  & 6.87
& 16.72
& 47.74
& 38.25
& 4.09
& 8.33
& 21.29
& 69.81
& 375.88
& 184.39
& 72.16
& 22
\\
        32  & 8.05
& 17.12
& 47.99
& 40.87
& 5.19
& 9.96
& 39.12
& 47.92
& 591.68
& 308.47
& 78.52
& 64.11
\\
        64  & 11.57
& 17.91
& 48.77& 43.23
& 7.59
& 14.04
& 48.84
& 61.05
& 809.96
& 437.47
& 125.79
& 100.65
\\
        128 & 20.53& 19.26& 50.08& 44.35& 12.85& 21.99
& 66.53& 83.69
& 956.61& 558.87
& 184.69& 146.83
\\
        \bottomrule
    \end{tabular}
    }
    \caption{Performance Comparison of 4 LLM Models on Different Batch Sizes for the \textbf{WikiText2 }dataset. All models operate in FP16 precision, except Deepseek-Qwen, which uses INT8. Experiments were conducted with Power Mode set to MaxN and sequence length (seqlen) equal to 96, comprising 32 input tokens and 64 generated tokens.}
    \label{tab:vary-batch-WT}
\end{table}

\subsubsection{LongBench Results}
\label{subsec:appendix-vary-batch-LB}

\begin{figure}[H]
    \centering
    \subfloat[Deepseek-Qwen-Distill\label{fig:deepQ-vary-batch-LB}]{
        \includegraphics[width=0.48\textwidth]{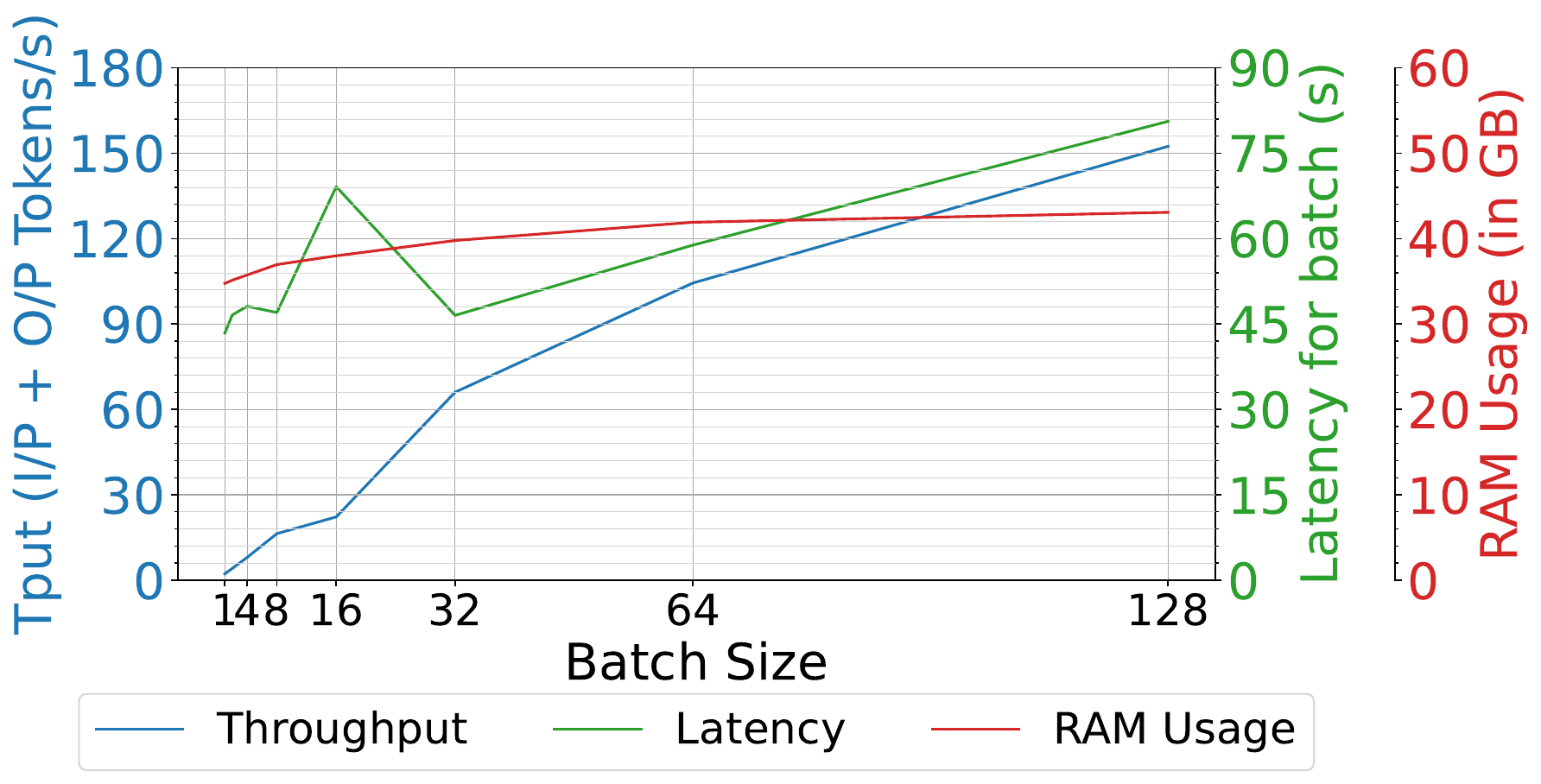}
    }
    \hfill
    \subfloat[Llama3.1-8B\label{fig:L3-vary-batch-LB}]{
        \includegraphics[width=0.48\textwidth]{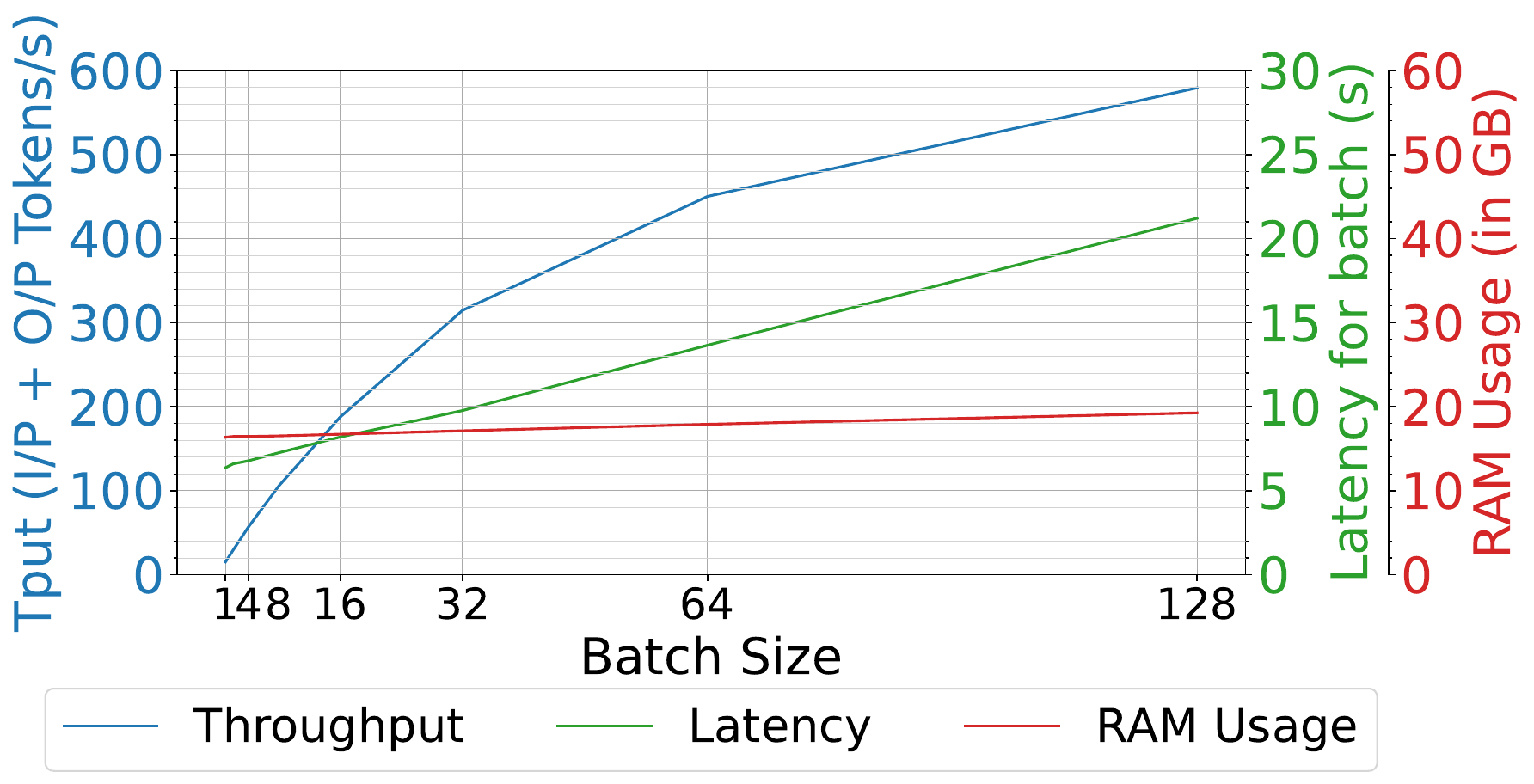}
    }
    
    \vspace{0.5cm} 
    
    \subfloat[Mistral-Base\label{fig:mistral-vary-batch-LB}]{
        \includegraphics[width=0.48\textwidth]{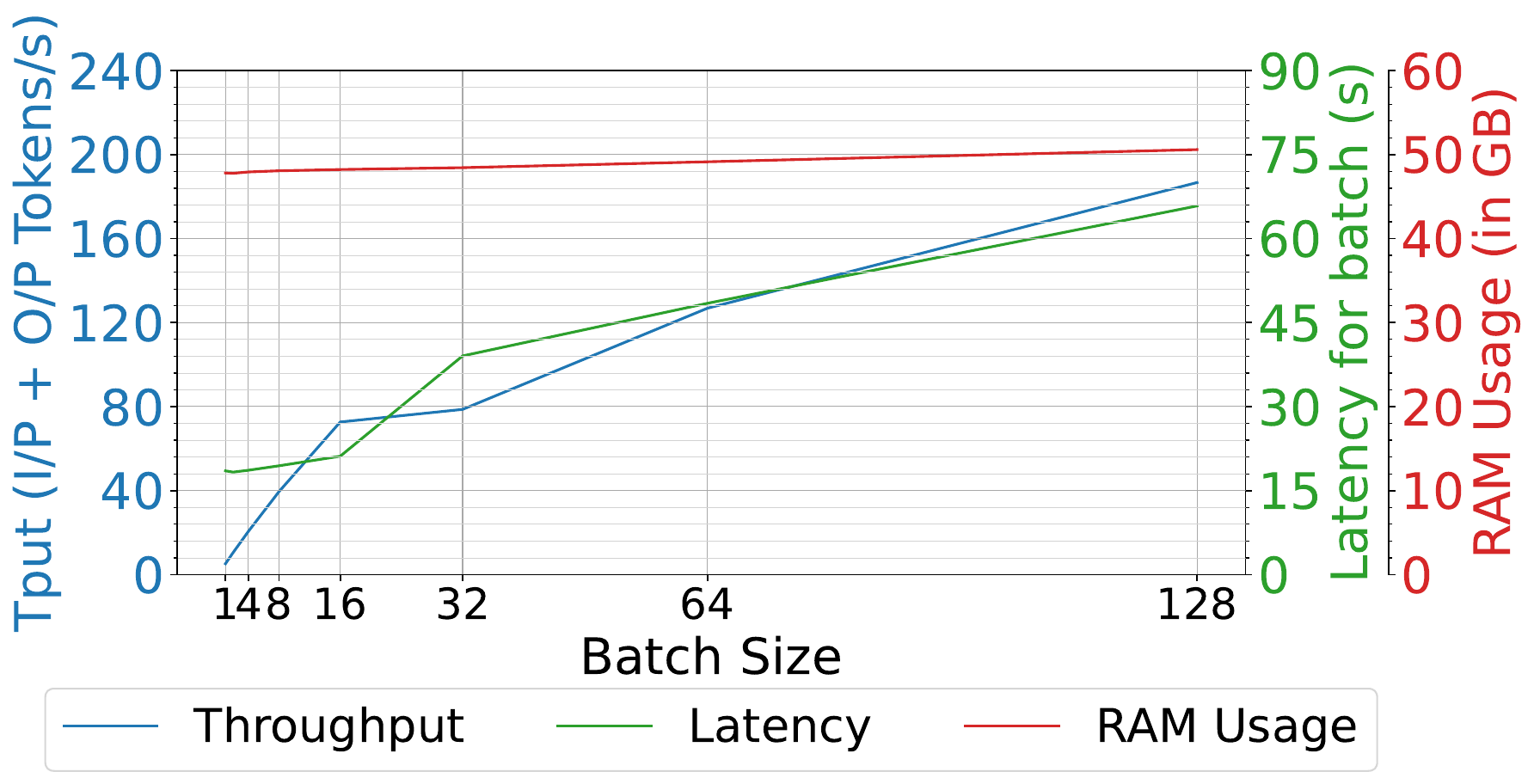}
    }
    \hfill
    \subfloat[MS-Phi2\label{fig:MS-Phi2-vary-batch-LB}]{
        \includegraphics[width=0.48\textwidth]{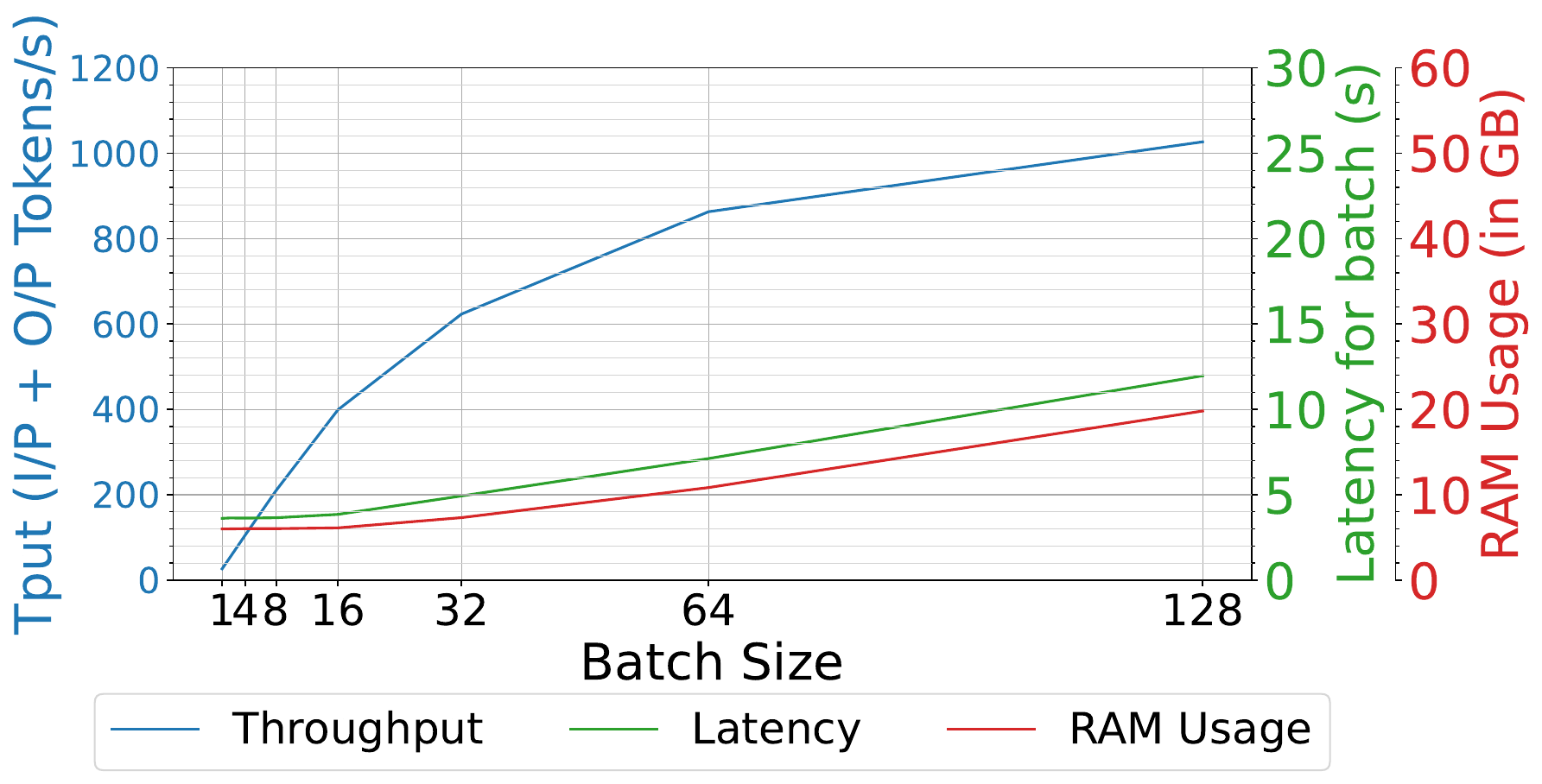}
    }
    
    \caption{Impact of varying batch sizes on different models evaluated using the LongBench dataset. All models operate in FP16 precision, except Deepseek-Qwen, which uses INT8. Experiments were conducted with Power Mode set to MaxN and sequence length (seqlen) equal to 96, comprising 32 input tokens and 64 generated tokens.}
    \label{fig:varying-batch-size-detailed-LB}
\end{figure}
Figure \ref{fig:varying-batch-size-detailed-LB} illustrates how increasing batch sizes affect memory consumption, end-to-end latency, and throughput for the \textit{LongBench} dataset, with detailed numerical results presented in Table \ref{tab:vary-batch-LB}. Compared to the \textit{WikiText2} dataset, the throughput variation remains within \textbf{$\approx10\%$}, despite an identical experimental setup. This variation could stem from dataset-specific and model-specific factors, minor fluctuations due to background workloads, or potential measurement inaccuracies.

\begin{table}[H]
    \centering
    \resizebox{\textwidth}{!}{ 
    \begin{tabular}{c|cccc|cccc|cccc}
        \toprule
        \textbf{Batch Size} & \multicolumn{4}{c|}{\textbf{RAM (GB)}} & \multicolumn{4}{c|}{\textbf{Latency (ms)}} & \multicolumn{4}{c}{\textbf{Throughput (tokens/s)}} \\
        & Phi2 & Llama3 & Mistral & DeepQ & Phi2 & Llama3 & Mistral & DeepQ & Phi2 & Llama3 & Mistral & DeepQ \\
        \midrule
        1   & 6.09
& 16.37
& 47.77
& 34.74
& 3.62
& 6.36
& 18.53
& 43.42
& 26.54
& 15.08
& 5.18
& 2.21
\\
        2   & 6.1
& 16.46
& 47.73
& 35.11
& 3.64
& 6.59
& 18.3
& 46.58
& 52.73
& 29.13
& 10.49
& 4.12
\\
        4   & 6.13
& 16.46
& 47.89
& 35.72
& 3.63
& 6.77
& 18.63
& 48.11
& 105.72
& 56.69
& 20.61
& 7.98
\\
        8   & 6.13
& 16.53
& 48.03
& 36.94
& 3.65
& 7.26
& 19.43
& 47.01
& 210.17
& 105.84
& 39.53
& 16.34
\\
        16  & 6.22
& 16.73
& 48.18
& 37.97
& 3.85
& 8.19
& 21.14
& 69.13
& 398.99
& 187.59
& 72.66
& 22.22
\\
        32  & 7.42
& 17.14
& 48.4
& 39.76
& 4.93
& 9.76
& 39.05
& 46.52
& 623.2
& 314.6
& 78.67
& 66.04
\\
        64  & 10.94
& 17.91
& 49.1& 41.9
& 7.12
& 13.65
& 48.44
& 58.86
& 863.01
& 450.12
& 126.83
& 104.39
\\
        128 & 19.91& 19.27& 50.55& 43.06& 11.97
& 21.21
& 65.83
& 80.61
& 1026.76
& 579.4
& 186.67
& 152.43
\\
        \bottomrule
    \end{tabular}
    }
    \caption{Performance Comparison of 4 LLM Models on different \textbf{Batch Sizes} for the \textbf{LongBench} dataset. All models operate in FP16 precision, except Deepseek-Qwen, which uses INT8. Experiments were conducted with Power Mode set to MaxN and sequence length (seqlen) equal to 96, comprising 32 input tokens and 64 generated tokens.}
    \label{tab:vary-batch-LB}
\end{table}

\subsection{Impact of Varying Sequence Length}

\begin{figure}[h!]
    \centering
    \subfloat[Deepseek-Qwen-Distill\label{fig:deepQ-vary-seqlen}]{
        \includegraphics[width=0.48\textwidth]{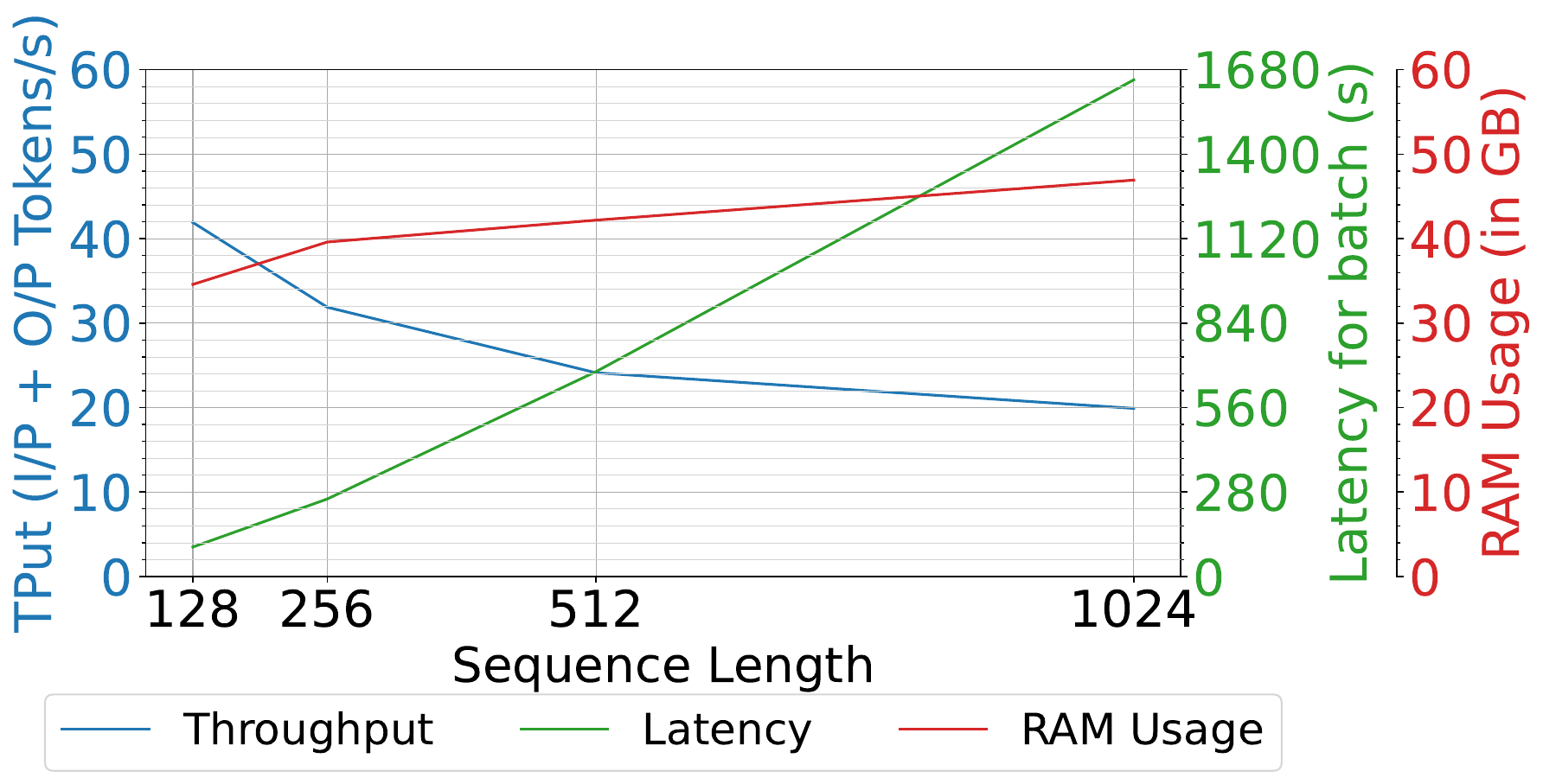}
    }
    \hfill
    \subfloat[Llama3.1-8B\label{fig:L3-vary-seqlen}]{
        \includegraphics[width=0.48\textwidth]{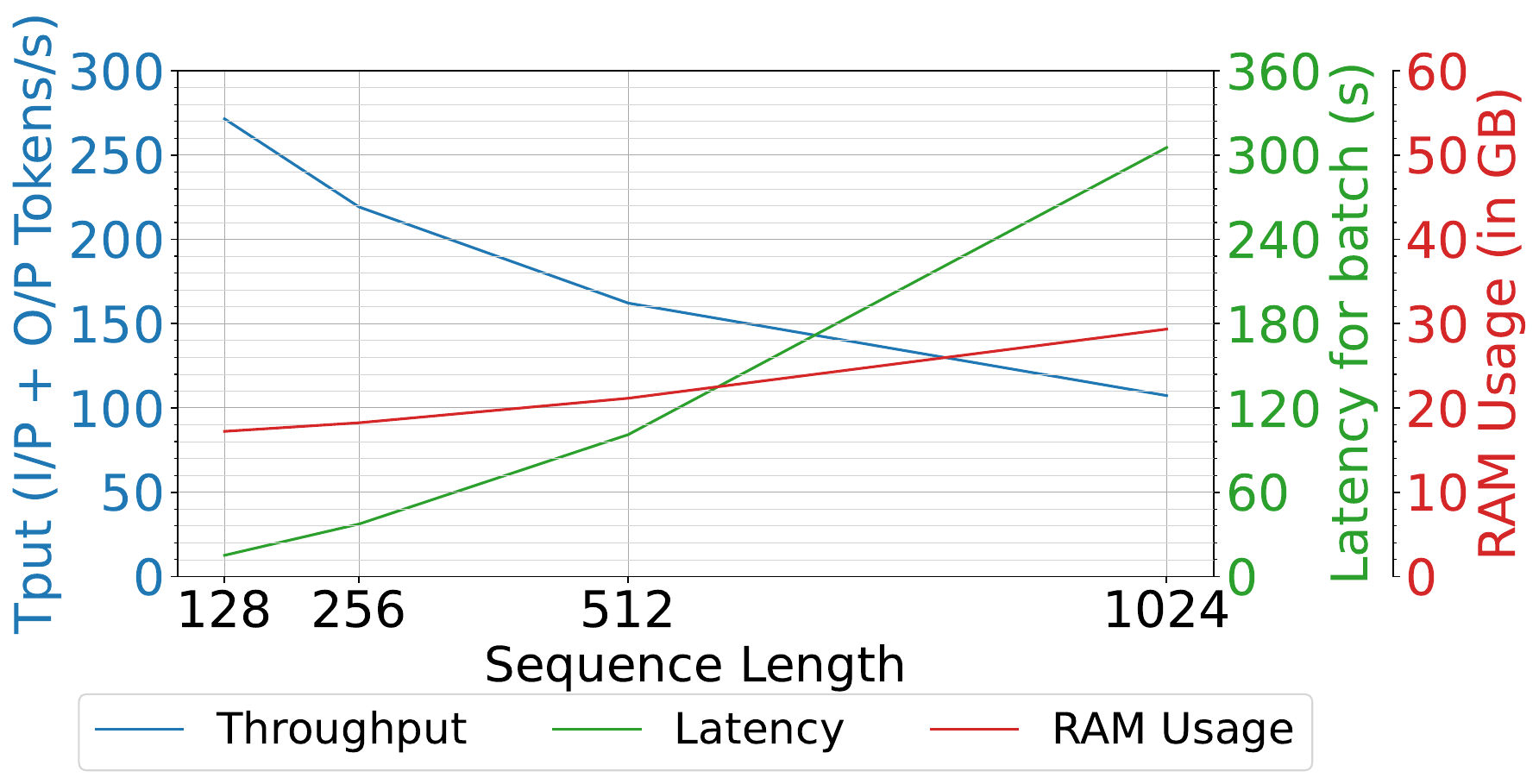}
    }
    
    \vspace{0.5cm} 
    
    \subfloat[Mistral-Base\label{fig:mistral-vary-seqlen}]{
        \includegraphics[width=0.48\textwidth]{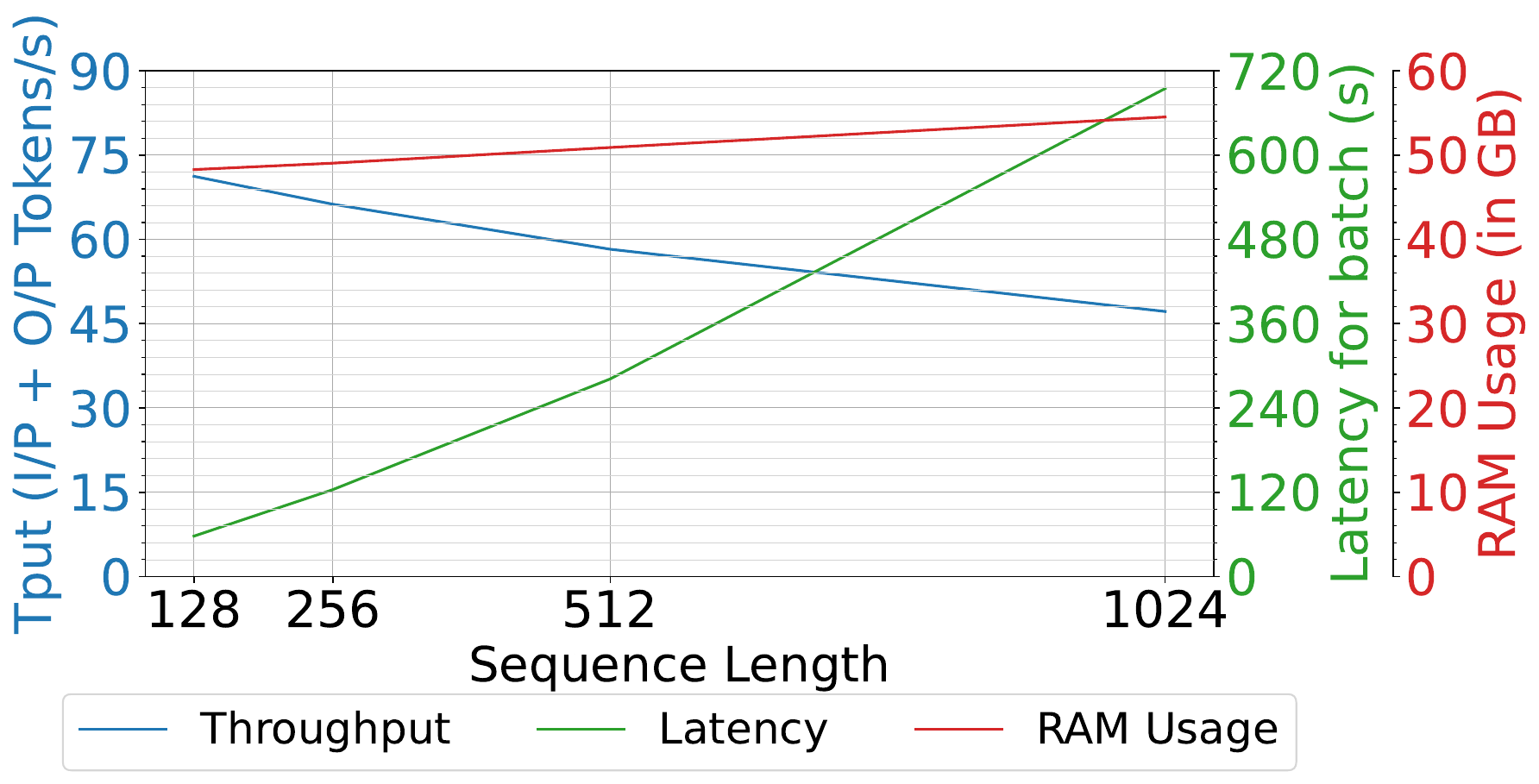}
    }
    \hfill
    \subfloat[MS-Phi2\label{fig:MS-Phi2-vary-seqlen}]{
        \includegraphics[width=0.48\textwidth]{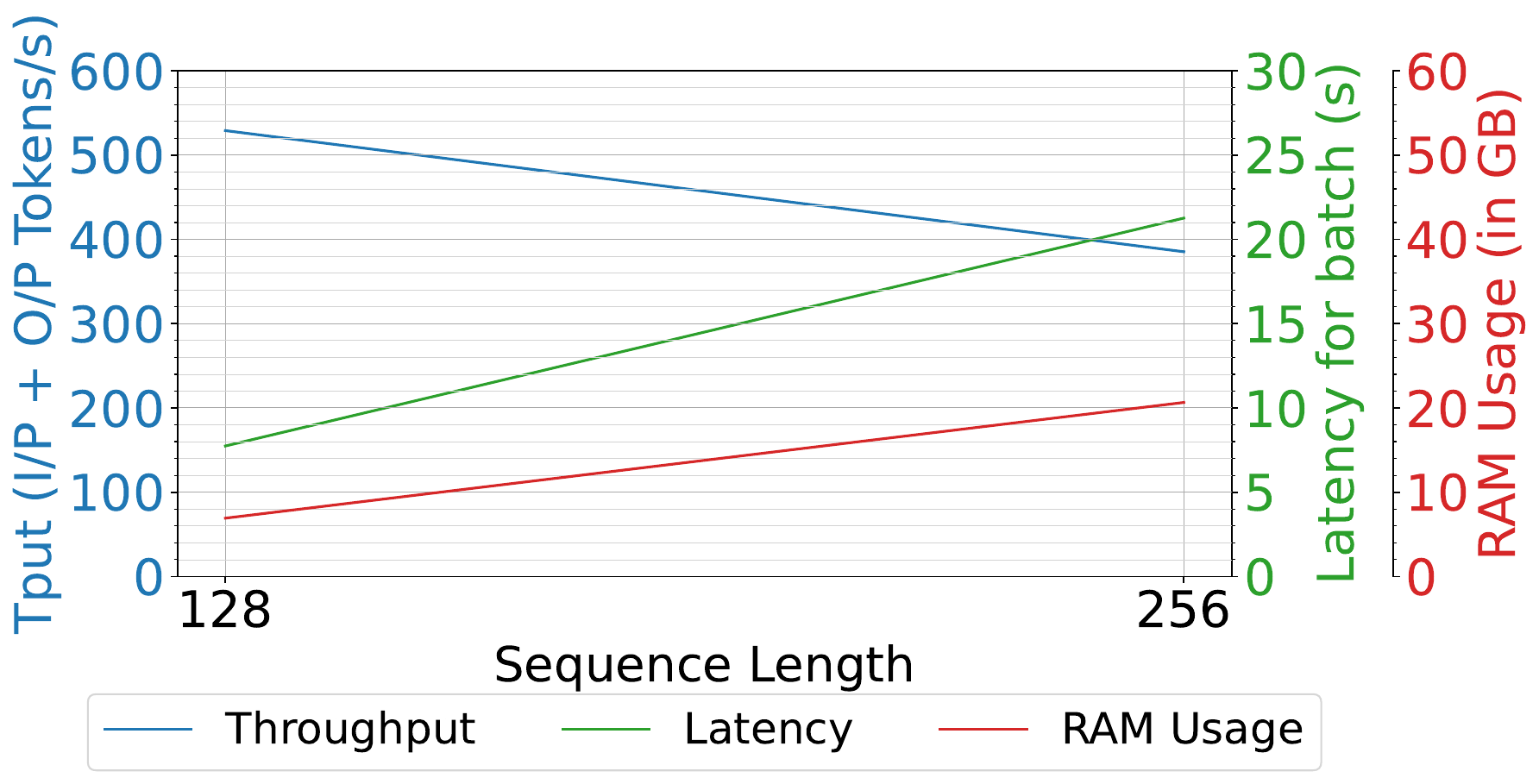}
    }
    
    \caption{Impact of varying sequence lengths on different models evaluated using the \textit{LongBench} dataset. All models operate in FP16 precision, except Deepseek-Qwen, which uses INT8. Experiments were conducted with Power Mode set to MaxN and a fixed batch size of 32.}
    \label{fig:vary-seqlen-detailed-LB}
\end{figure}
\begin{table}[h!]
    \centering
    \resizebox{\textwidth}{!}{ 
    \begin{tabular}{c|cccc|cccc|cccc}
        \toprule
        \textbf{Sequence Length}& \multicolumn{4}{c|}{\textbf{RAM (GB)}} & \multicolumn{4}{c|}{\textbf{Latency (ms)}} & \multicolumn{4}{c}{\textbf{Throughput (tokens/s)}} \\
        & Phi2 & Llama3 & Mistral & DeepQ & Phi2 & Llama3 & Mistral & DeepQ & Phi2 & Llama3 & Mistral & DeepQ \\
        \midrule
        128& 6.97
& 17.24
& 48.24
& 34.56
& 7.74
& 15.09
& 57.51
& 97.72
& 529.04
& 271.5
& 71.22
& 41.91
\\
        256& 20.7& 18.26
& 49
& 39.58
& 21.26
& 37.37
& 123.64
& 257.02
& 385.32
& 219.21
& 66.26
& 31.88
\\
        512& OOM& 21.17
& 50.86
& 42.17
& OOM& 101.02
& 281.3
& 679.31
& OOM& 162.18
& 58.24
& 24.12
\\
        1024& OOM& 29.37& 54.48& 46.91& OOM& 305.36
& 694.74
& 1646.36
& OOM& 107.31
& 47.17
& 19.9
\\
        \bottomrule
    \end{tabular}
    }
    \caption{Performance Comparison of 4 LLM Models on varying\textbf{ Sequence Length }for the \textbf{LongBench} dataset. Note that \textbf{DeepQ} is being run in \textbf{INT8} precision while all other models in \textbf{FP16}.}
    \label{tab:vary-seqlen-LB}
\end{table}

This section expands upon (\S\ref{subsec:impact-of-seqlen}). In our experiments, we vary the overall sequence length — the sum of the input prompt tokens and the generated output tokens. Specifically, we evaluate sequence lengths defined as A = B + C, where B is the number of input tokens and C is the number of generated tokens. The configurations we consider are: $128$ tokens ($32$ input + $96$ output), $256$ tokens ($64$ input + $192$ output), $512$ tokens ($128$ input + $384$ output), and 1024 tokens ($256$ input + $768$ output). Because our sequence length configurations are primarily dominated by the number of output tokens (which significantly outweighs the input tokens), we consistently observe a decrease in throughput and an increase in end-to-end latency. This performance drop occurs because the decoding phase—the token generation stage of LLM inference—is memory-bound \cite{patel2024splitwiseefficientgenerativellm}. In this phase, memory constraints slow down token production, directly impacting overall throughput and latency. Figure \ref{fig:vary-seqlen-detailed-LB} illustrates the impact of varying sequence length on the various models being evaluated in this study.

\subsubsection{WikiText2 Results}

\begin{figure}[h]
    \centering
    \subfloat[Deepseek-Qwen-Distill\label{fig:deepQ-vary-seqlen-WT}]{
        \includegraphics[width=0.48\textwidth]{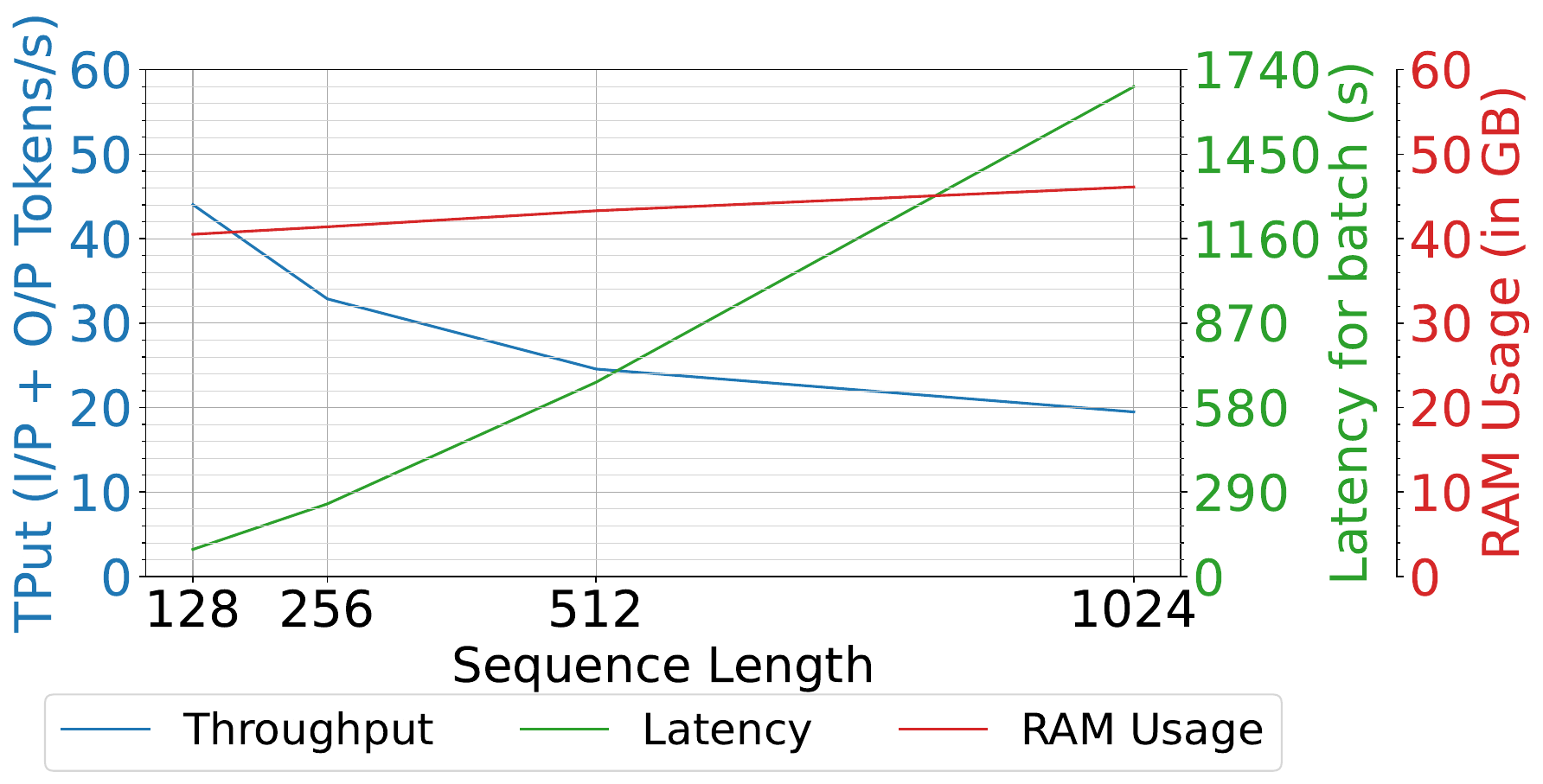}
    }
    \hfill
    \subfloat[Llama3.1-8B\label{fig:L3-vary-seqlen-WT}]{
        \includegraphics[width=0.48\textwidth]{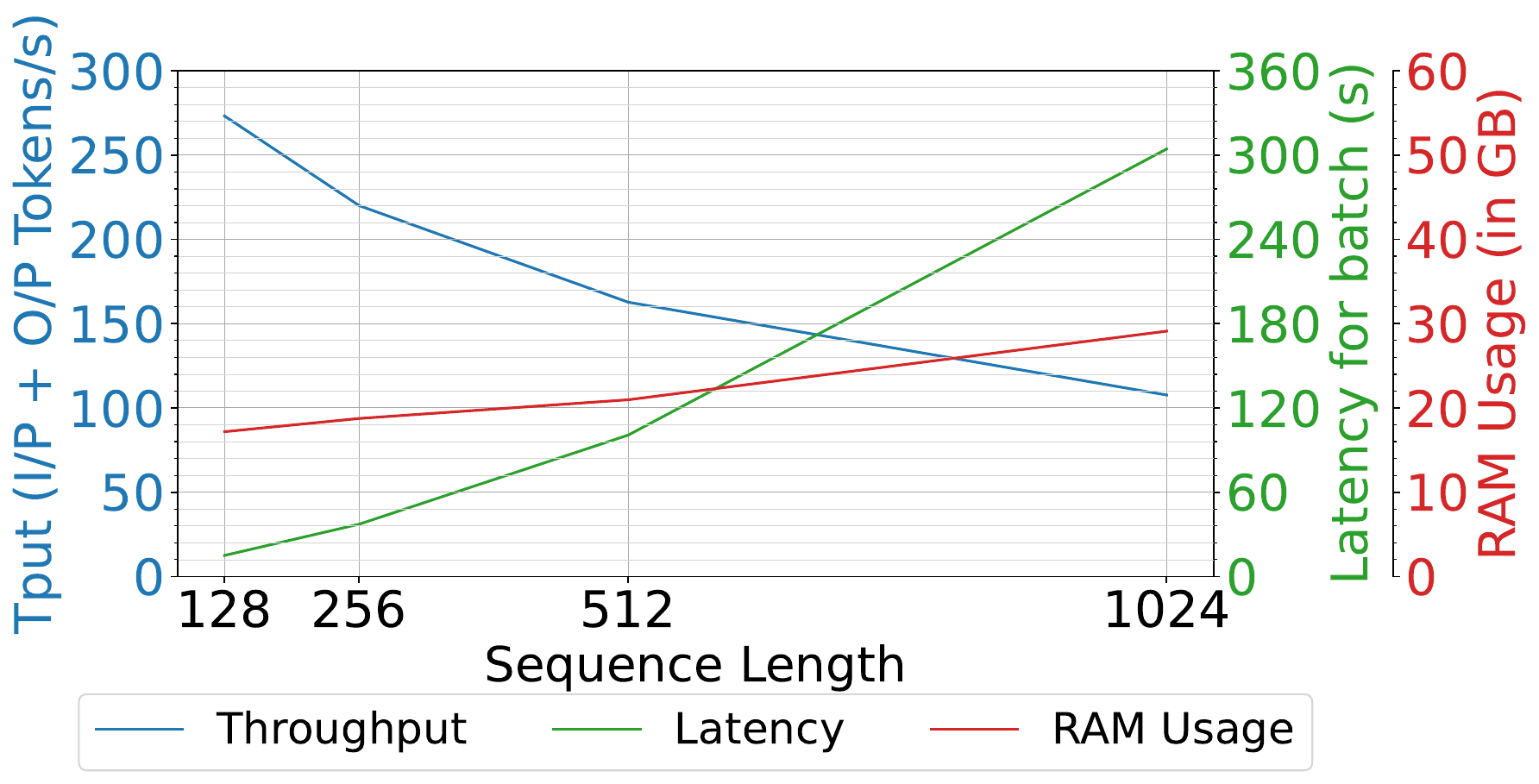}
    }
    
    \vspace{0.5cm} 
    
    \subfloat[Mistral-Base\label{fig:mistral-vary-seqlen-WT}]{
        \includegraphics[width=0.48\textwidth]{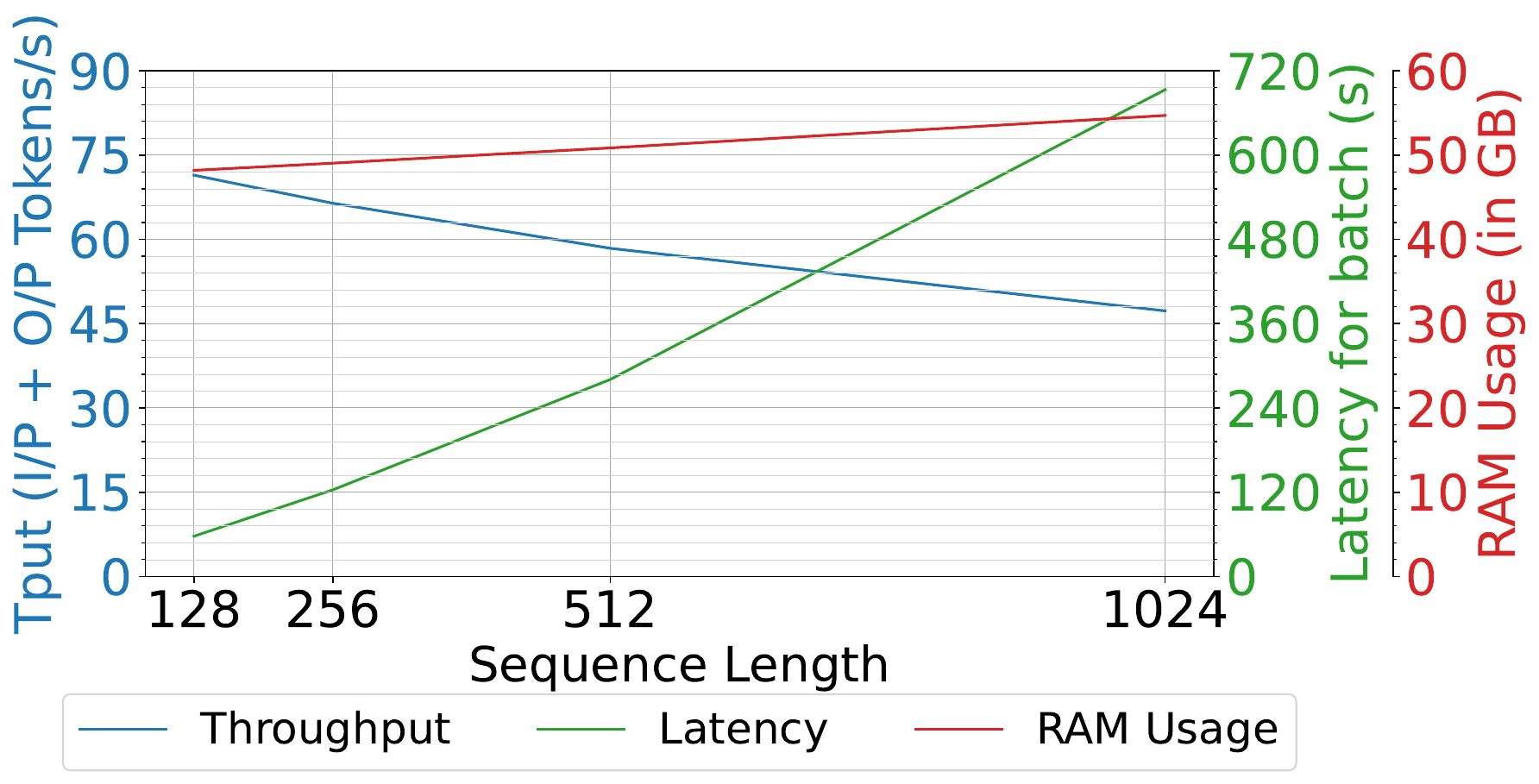}
    }
    \hfill
    \subfloat[MS-Phi2\label{fig:MS-Phi2-vary-seqlen-WT}]{
        \includegraphics[width=0.48\textwidth]{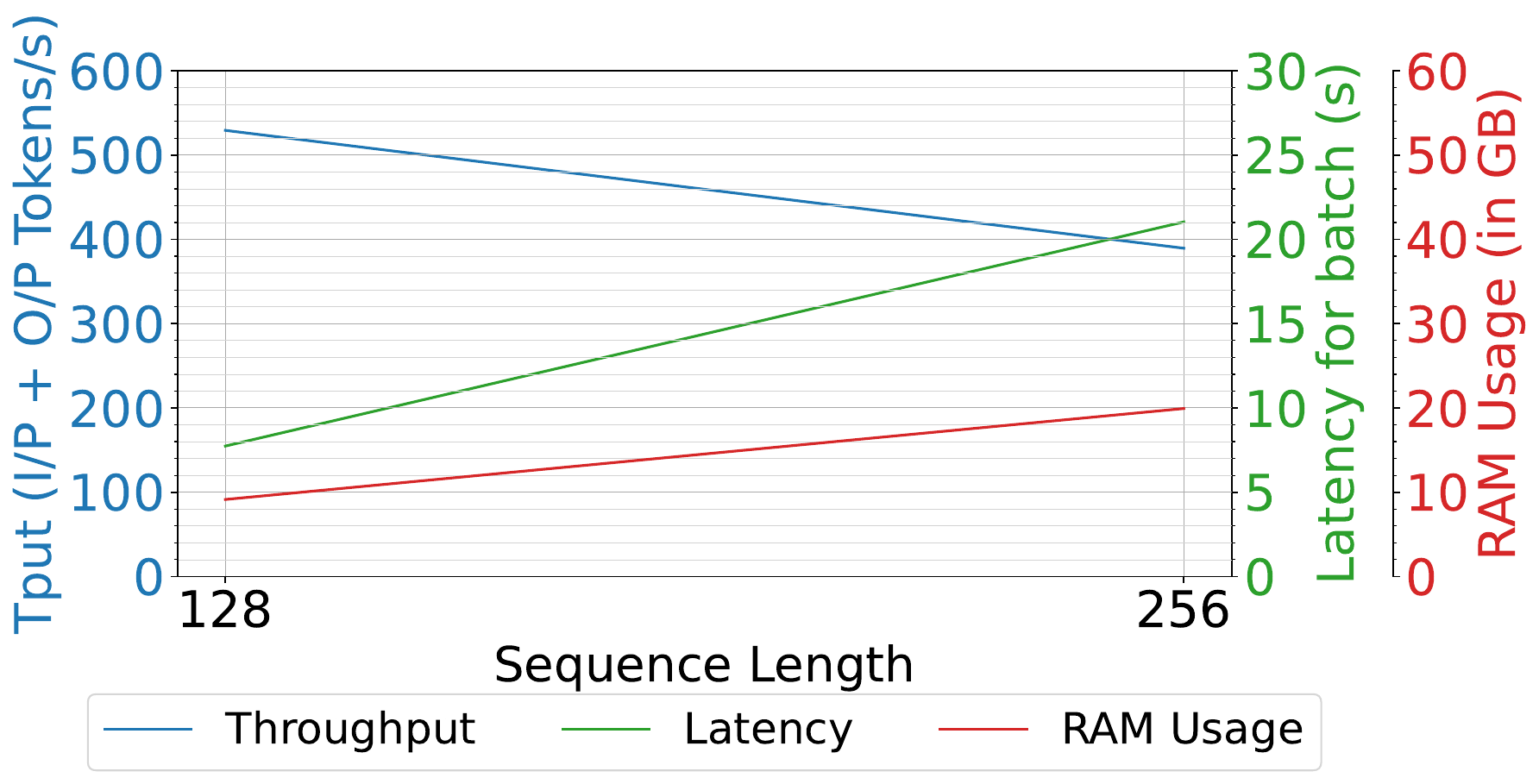}
    }
    
    \caption{Impact of varying sequence lengths on different models evaluated using the \textit{WikiText2} dataset. All models operate in FP16 precision, except Deepseek-Qwen, which uses INT8. Experiments were conducted with Power Mode set to MaxN and a fixed batch size of 32.}
    \label{fig:vary-seqlen-detailed-WT}
\end{figure}

While we reported numbers for the \textit{LongBench} dataset in the paper (\S\ref{subsec:impact-of-seqlen}), here we present the results for the \textit{WikiText2} dataset. Figure \ref{fig:vary-seqlen-detailed-WT} shows the impact of varying sequence length on the inference throughput, the end-to-end latency and the overall shared (CPU + GPU) memory utilization.

\begin{table}[h!]
    \centering
    \resizebox{\textwidth}{!}{ 
    \begin{tabular}{c|cccc|cccc|cccc}
        \toprule
        \textbf{Sequence Length}& \multicolumn{4}{c|}{\textbf{RAM (GB)}} & \multicolumn{4}{c|}{\textbf{Latency (ms)}} & \multicolumn{4}{c}{\textbf{Throughput (tokens/s)}} \\
        & Phi2 & Llama3 & Mistral & DeepQ & Phi2 & Llama3 & Mistral & DeepQ & Phi2 & Llama3 & Mistral & DeepQ \\
        \midrule
        128& 9.19
& 17.2
& 48.15
& 40.49
& 7.74
& 14.99
& 57.35
& 93.04
& 529.31
& 273.18
& 71.42
& 44.03
\\
        256& 19.98& 18.77
& 49
& 41.38
& 21.03
& 37.23
& 123.31
& 249.24
& 389.48
& 220.02
& 66.43
& 32.87
\\
        512& OOM& 20.99
& 50.81
& 43.28
& OOM& 100.69
& 280.48
& 667.08
& OOM& 162.71
& 58.41
& 24.56
\\
        1024& OOM& 29.13& 54.66& 46.1& OOM& 304.33
& 693.13
& 1681.75
& OOM& 107.67
& 47.28
& 19.48
\\
        \bottomrule
    \end{tabular}
    }
    \caption{Performance Comparison of 4 LLM Models on varying\textbf{ Sequence Length }for the \textbf{WikiText2} dataset. Note that \textbf{DeepQ} is being run in \textbf{INT8} precision while all other models in \textbf{FP16}.}
    \label{tab:vary-seqlen-WT}
\end{table}

\subsection{Impact of quantization}

We reported the impact of quantization on power and energy consumption for the Llama3-8B model in the paper (\S\ref{subsec:impact-of-quantization}). Here, in Fig. \ref{fig:power-energy-quant-WT} we expand those results for all other models evaluated in the study. The INT8 configuration generally offers the most favorable efficiency, though relative gains vary by model and batch size:
\begin{enumerate}
    \item Llama3-8b: \begin{itemize}
        \item Power Consumption: INT8 shows up to 44\% less power usage than FP16 (with a median reduction of 39\%), though at peak batch sizes (e.g., batch size 16) savings drop to about 26\%. When compared with INT4, INT8’s power savings range between 20\% and 43\% (median 32\%).
        \item Energy Consumption: FP16 exhibits the lowest energy consumption overall. Specifically, FP16 consumes about 23\% less energy than INT8 (with near convergence at larger batch sizes) and achieves a median savings of approximately 78\%—reaching up to 82\% in the best cases—compared to INT4.
    \end{itemize}

\begin{figure}[H]
    \centering
    \subfloat[Deepseek-Qwen-Distill\label{fig:deepQ-quant-power}]{
        \includegraphics[width=0.48\textwidth]{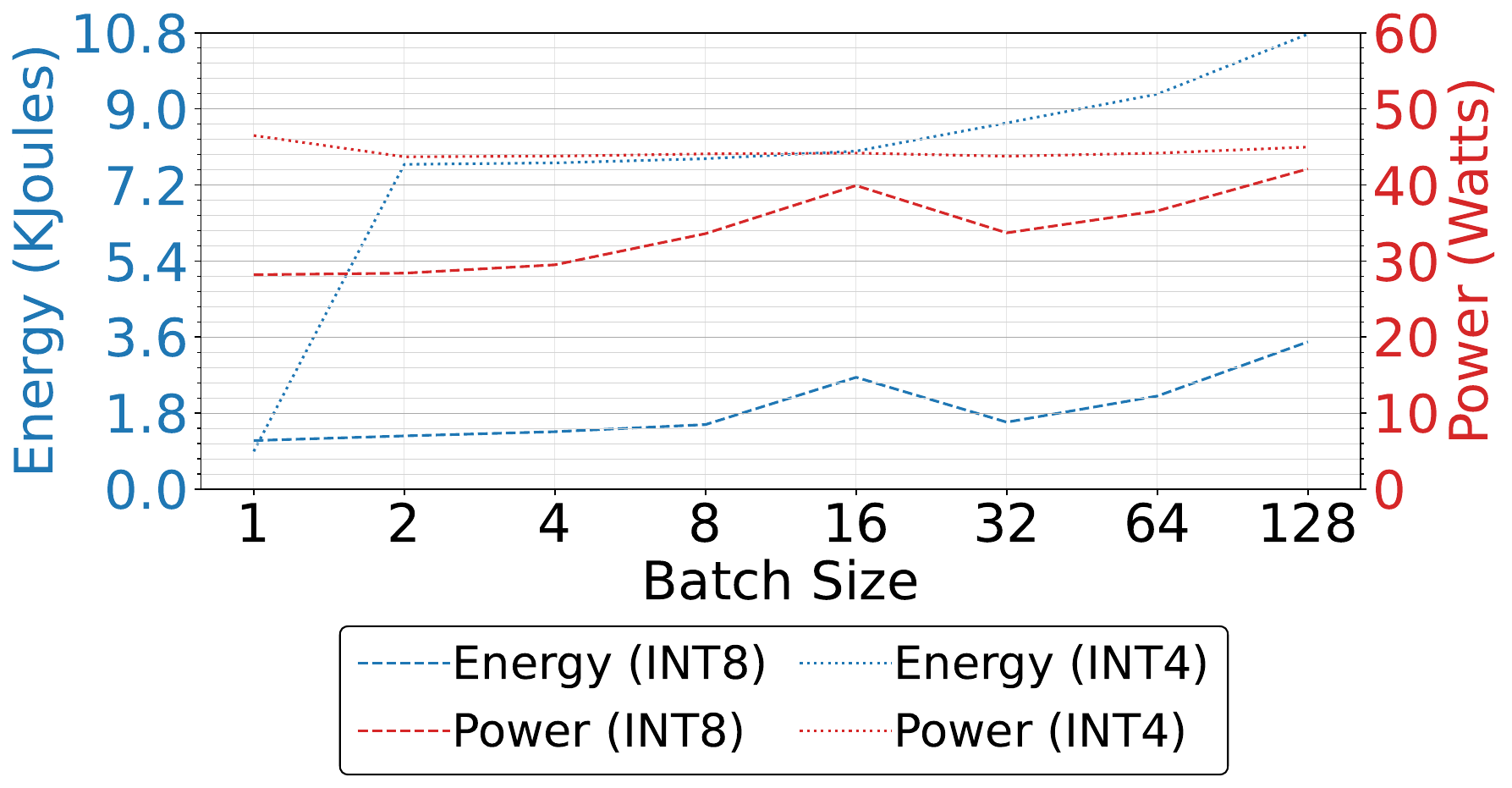}
    }
    \hfill
    \subfloat[Llama3.1-8B\label{fig:L3-quant-power}]{
        \includegraphics[width=0.48\textwidth]{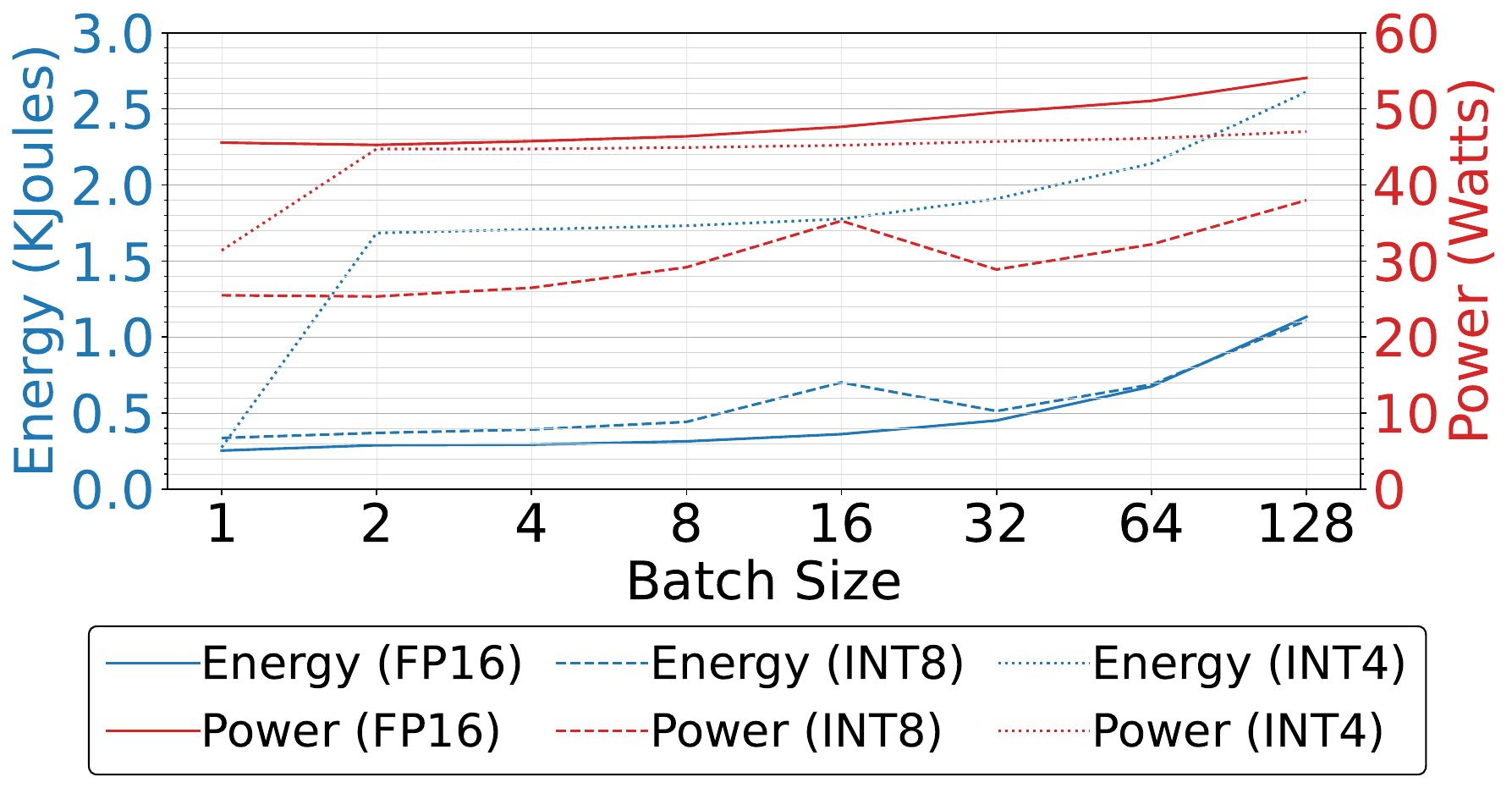}
    }
    
    \vspace{0.5cm}  
    
    \subfloat[Mistral-Base\label{fig:mistral-quant-power}]{
        \includegraphics[width=0.48\textwidth]{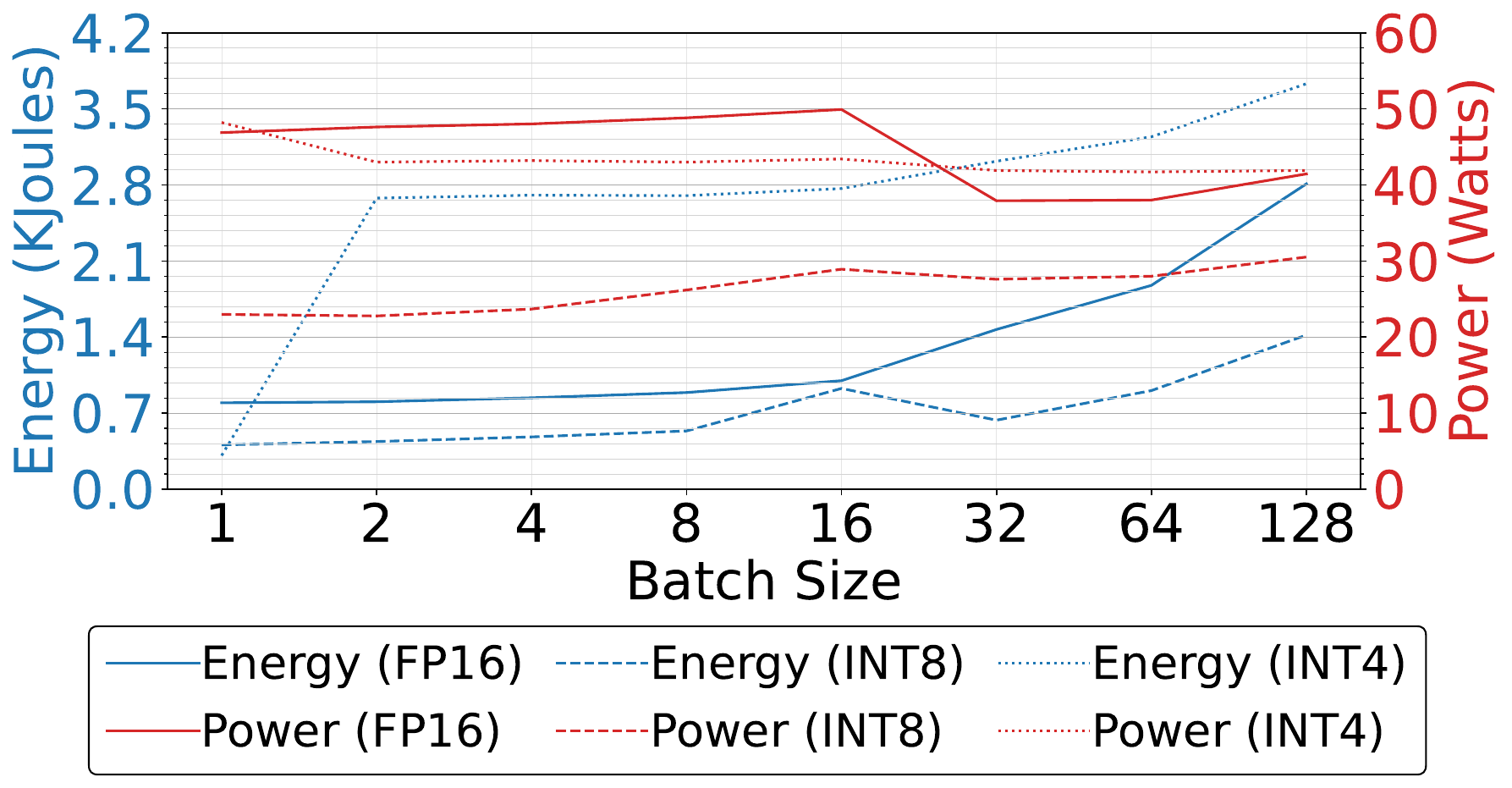}
    }
    \hfill
    \subfloat[MS-Phi2\label{fig:MS-Phi2-quant-power}]{
        \includegraphics[width=0.48\textwidth]{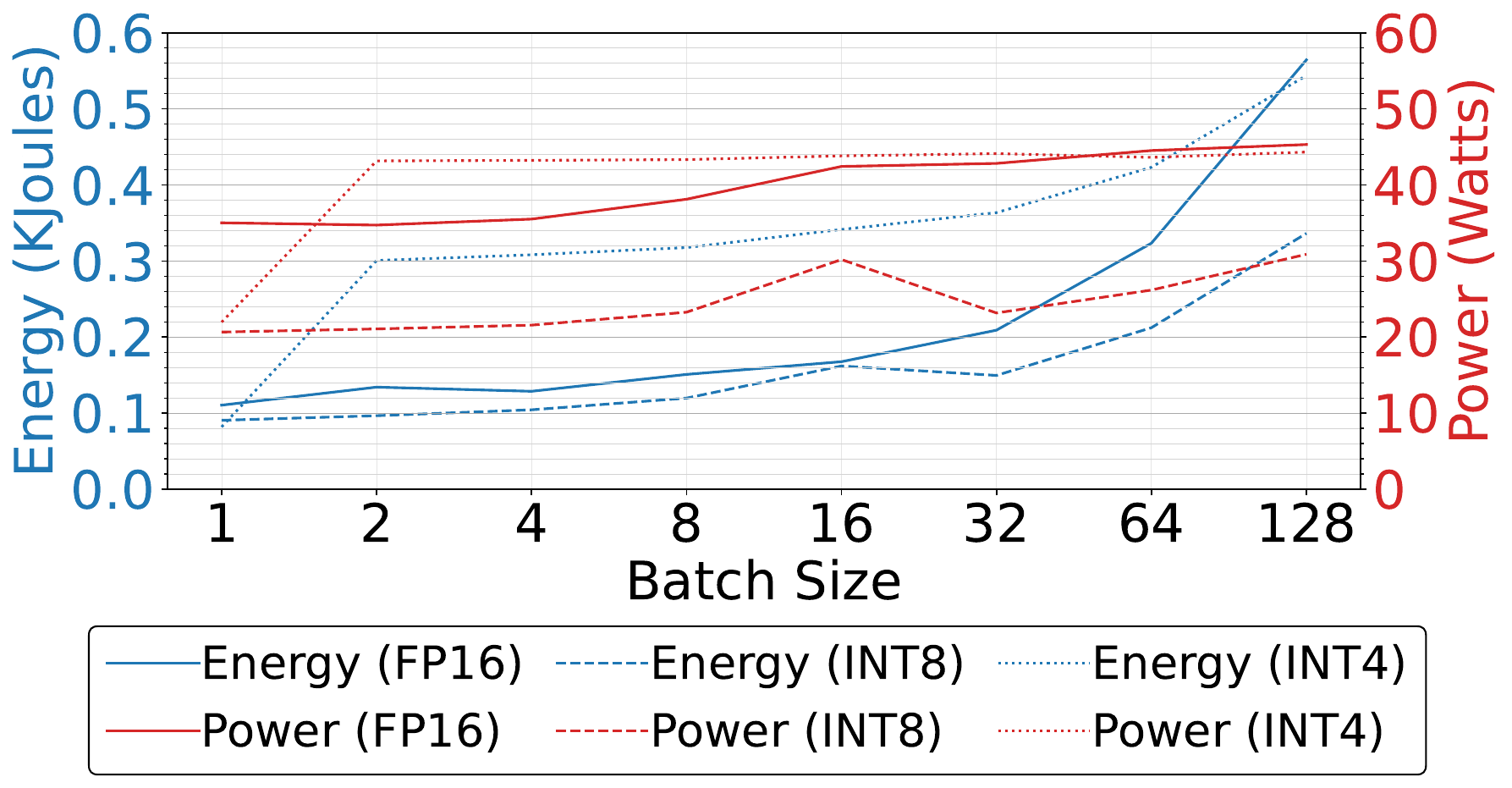}
    }
    
    \caption{Impact of varying quantization on Power and Energy Consumption, across increasing batch size. The device is operating on MaxN power mode, with sequence length = $96$ ($32$ input + $64$ output tokens)}
    \label{fig:power-energy-quant-WT}
\end{figure}

    \item Mistral-Base: \begin{itemize}
        \item Power Consumption: INT8 provides the lowest power draw, saving around 50\% relative to FP16 for smaller batch sizes, though savings reduce to roughly 26\% for larger batches. Against INT4, INT8 consistently yields over 27\% power savings.
        \item Energy Consumption: In this model, INT8 is also superior in energy efficiency, offering a median reduction of 47\% compared to FP16 and 75\% relative to INT4. Notably, INT4’s energy consumption is about 57\% higher than FP16 from batch sizes of 2 onward.
    \end{itemize}

    \item MS Phi-2: \begin{itemize}
        \item Power Consumption: INT8 continues to deliver significant power reductions, with a median decrease of 39\% compared to FP16 and 43\% compared to INT4.
        \item Energy Consumption: Similarly, INT8 achieves lower energy usage, with a median reduction of 24\% compared to FP16 and 55\% compared to INT4.
    \end{itemize}

    \item Deepseek-Qwen: \begin{itemize}
        \item Due to memory constraints, FP16 could not be executed.
        \item Power \& Energy Consumption: Among the available configurations, INT8 records a median power reduction of 23\% and exhibits a substantial energy consumption advantage, with a median savings of 78\% relative to INT4.
    \end{itemize}
\end{enumerate}

\begin{figure}[H]
    \centering
    \subfloat[Deepseek-Qwen-Distill\label{fig:deepQ-quant-thr-lat-mem}]{
        \includegraphics[width=0.48\textwidth]{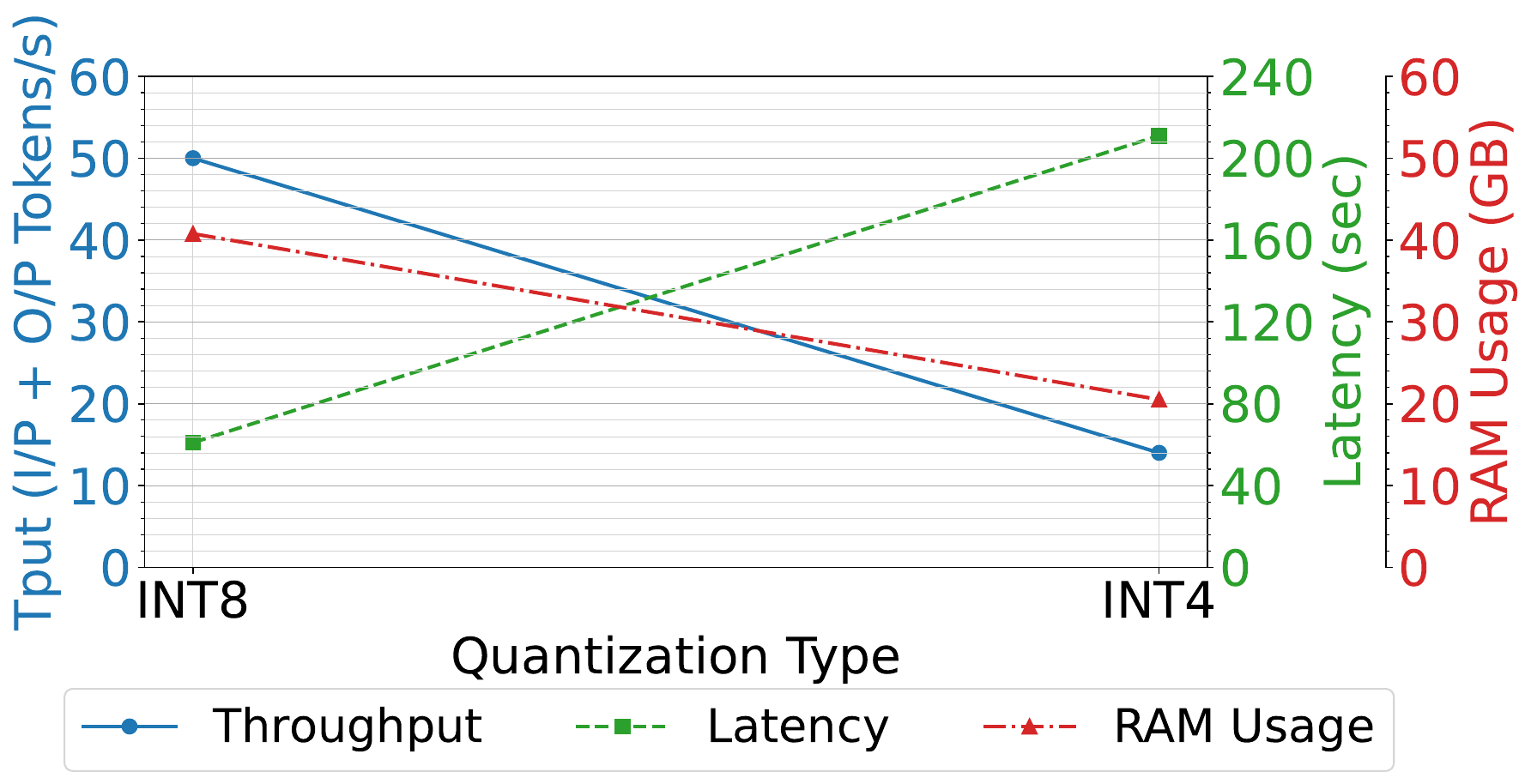}
    }
    \hfill
    \subfloat[Llama3.1-8B\label{fig:L3-quant-thr-lat-mem}]{
        \includegraphics[width=0.48\textwidth]{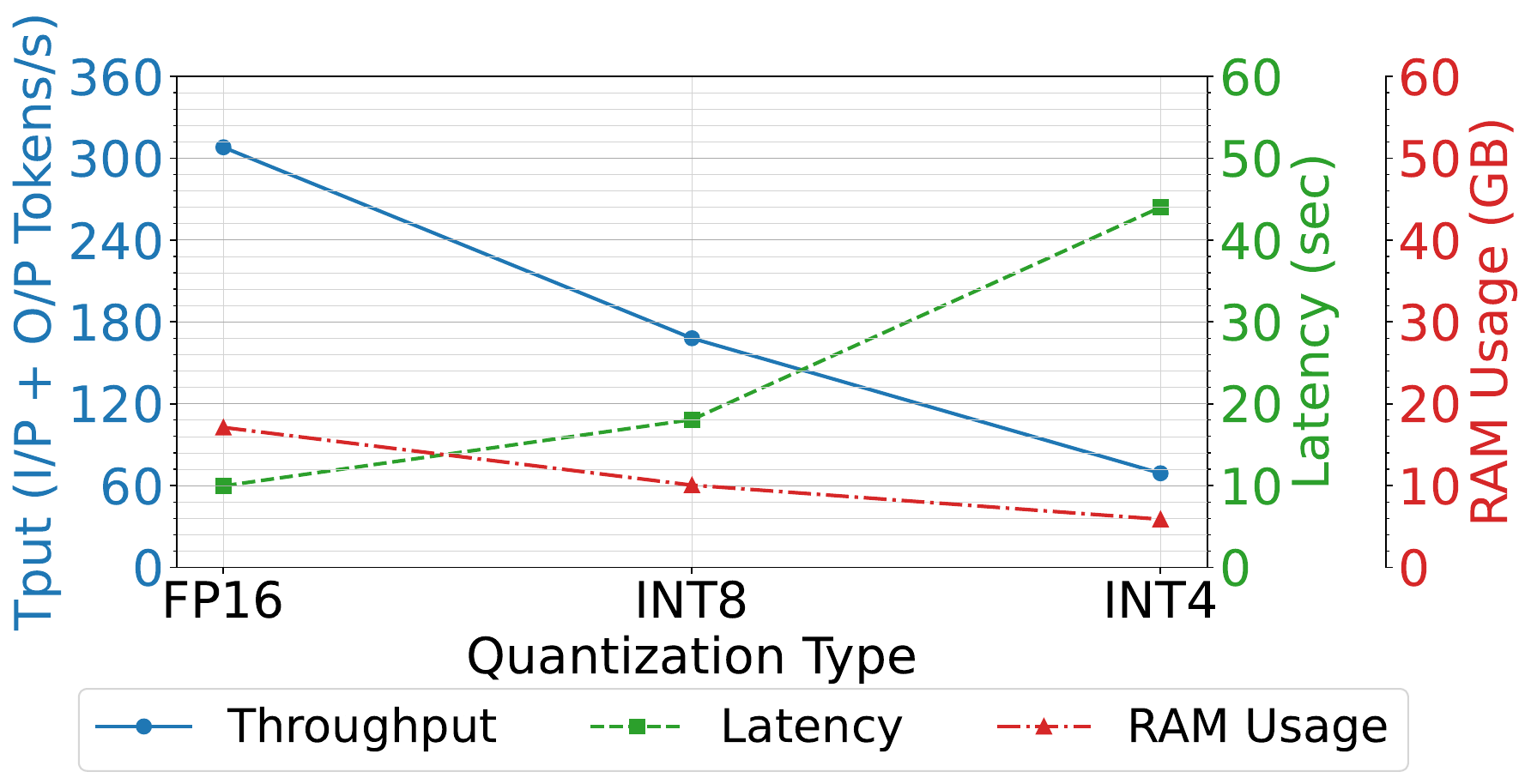}
    }
    
    \subfloat[Mistral-Base\label{fig:mistral-quant-thr-lat-mem}]{
        \includegraphics[width=0.48\textwidth]{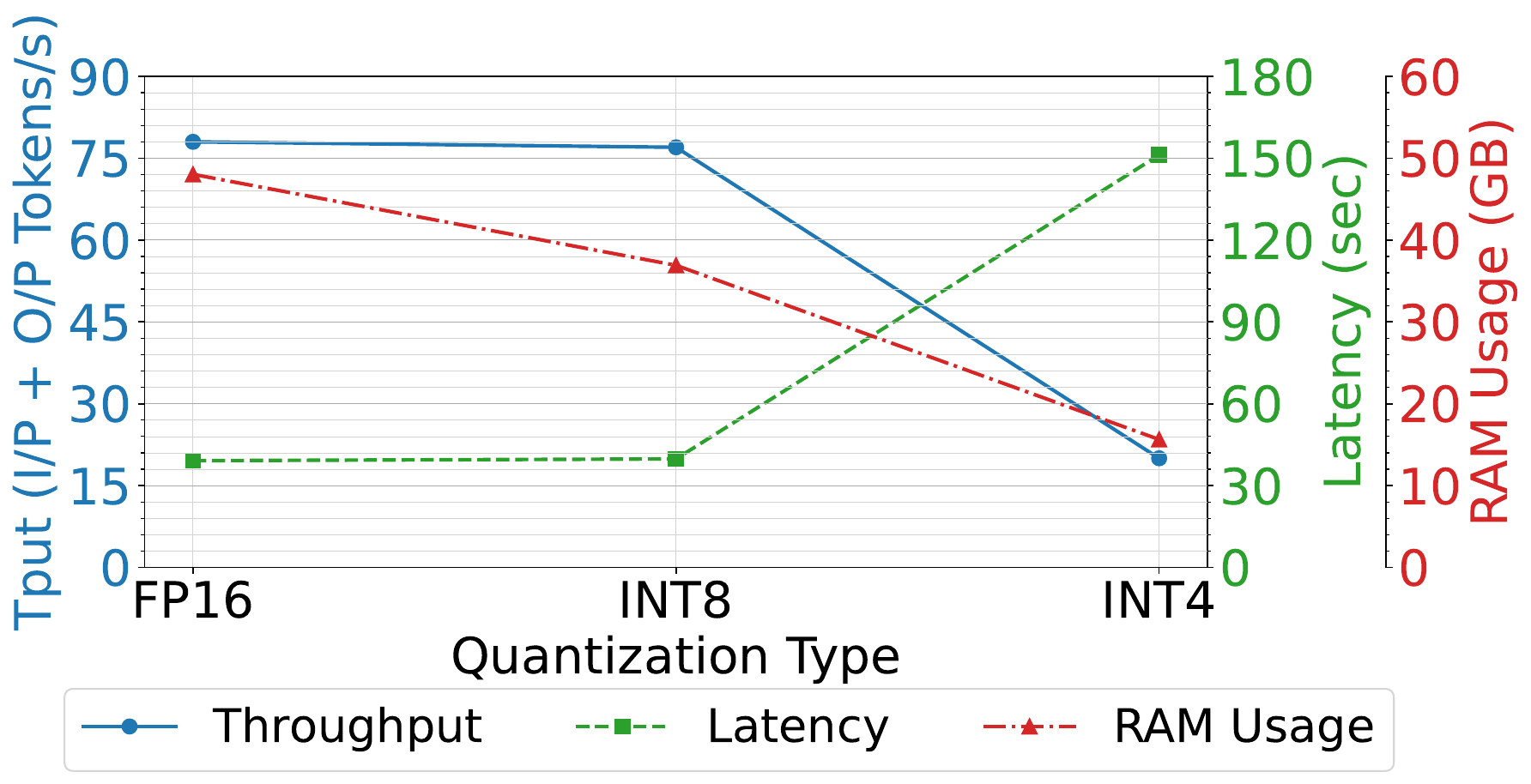}
    }
    \hfill
    \subfloat[MS-Phi2\label{fig:MS-Phi2-quant-thr-lat-mem}]{
        \includegraphics[width=0.48\textwidth]{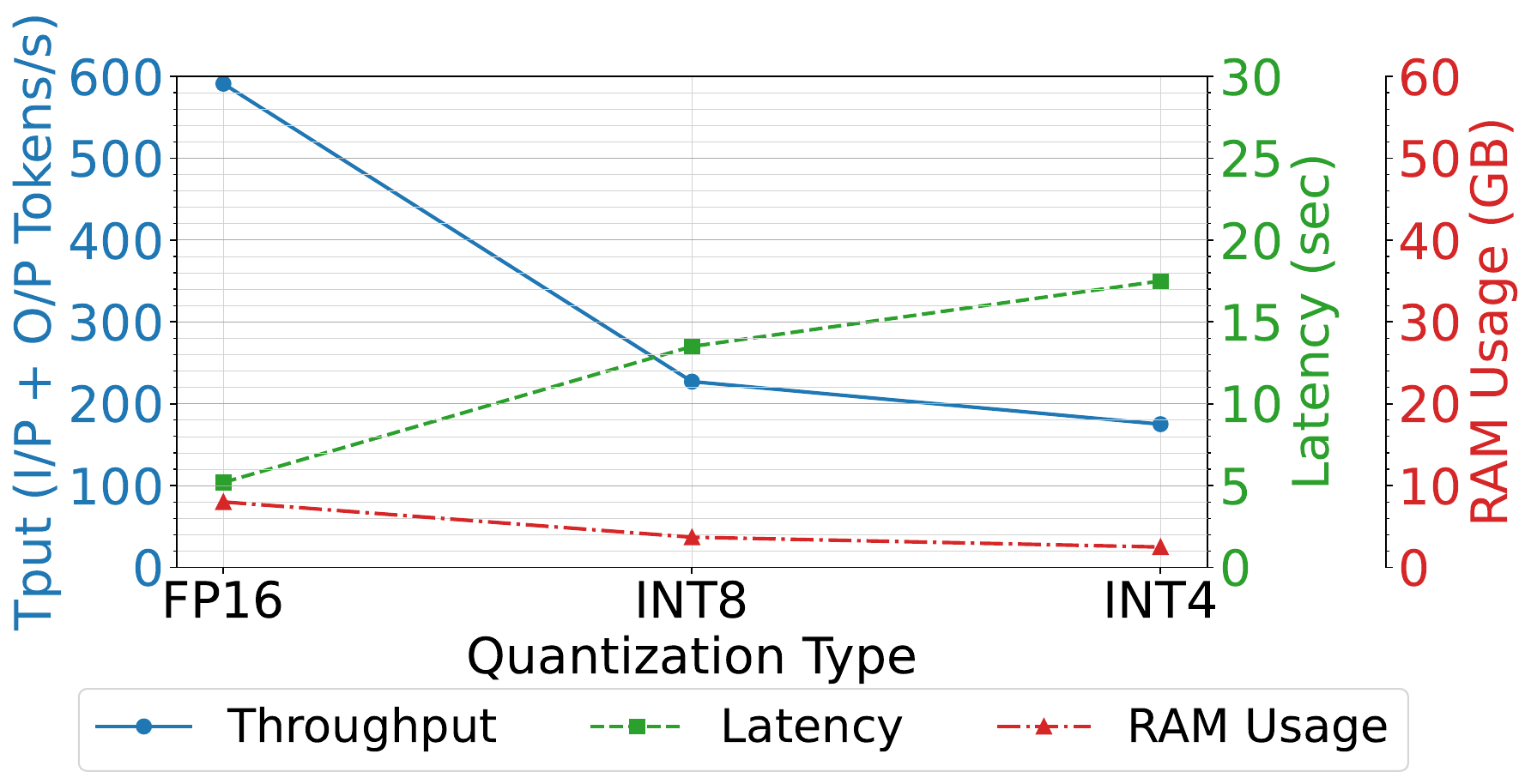}
    }
    
    \caption{Impact of varying quantization on model throughput, end-to-end token generation latency, and total shared memory consumption. The device is operating on MaxN power mode, batch size is fixed to $32$ with sequence length = $96$ ($32$ input + $64$ output tokens)}
    \label{fig:thr-lat-mem-quant-WT}
\end{figure}

Transitioning from FP16 to lower precisions like INT8 and INT4 consistently reduces the memory footprint across all models. However, this reduction in memory utilization is accompanied by increased end-to-end latency for token generation and a decrease in throughput (Fig. \ref{fig:thr-lat-mem-quant-WT}). These performance trade-offs are highlighted in literature \cite{NEURIPS2022_c3ba4962}, which notes that while INT8 matrix multiplication can accelerate inference for models with large model and hidden dimensions, overall benefits vary by model. Notably, the Mistral model exhibits minimal throughput degradation and maintains latency similar to FP16 when using INT8, emphasizing that quantization effects are model-dependent.

\end{document}